\documentclass[aps,pra,twocolumn,superscriptaddress]{revtex4-2}

\usepackage[pages=all, color=black, position={current page.south}, placement=bottom, scale=1, opacity=1, vshift=5mm]{background}
\SetBgContents{
	\tt 
}      

\usepackage{amsmath}
\usepackage{amsthm}
\usepackage{amssymb}
\usepackage{times}
\usepackage{braket}
\usepackage{ulem}

\usepackage[utf8]{inputenc}
\usepackage{hyperref}

\usepackage{graphicx, color}
\graphicspath{{fig/}}

\newcommand{\ptr}[2]{\mathrm{tr}_{#2}\left \{ #1 \right\}}

\definecolor{harvardcrimson}{rgb}{0.79, 0.0, 0.09}

\definecolor{chromeyellow}{rgb}{1.0, 0.65, 0.0}

\usepackage{algorithm, algpseudocode} % use algorithm and algorithmicx for typesetting algorithms
\usepackage{mathrsfs} % for \mathscr command
\usepackage{lipsum}

\begin{document}

%\title{Quantum photo-thermodynamics on a programmable quantum photonic processor}

\title{Quantum simulation of thermodynamics in an integrated quantum photonic processor}

\author{F.~H.~B.~Somhorst}
\author{R.~van der Meer}
\author{M.~Correa Anguita}
\affiliation{MESA+ Institute for Nanotechnology, University of Twente, P.~O.~box 217, 7500 AE Enschede, The Netherlands} 
\author{R.~Schadow}
\affiliation{Dahlem Center for Complex Quantum Systems, Freie Universit{\"a}t Berlin, 14195 Berlin, Germany}

\author{H.~J.~Snijders}
\author{M.~de Goede}
\author{B.~Kassenberg}
\author{P.~Venderbosch}
\author{C.~Taballione}
\author{J.~P.~Epping}
\author{H.~H.~van den Vlekkert}
\affiliation{QuiX Quantum B.V., Hengelosestraat 500, 7521 AN Enschede, the Netherlands}

\author{J. Timmerhuis}
\affiliation{MESA+ Institute for Nanotechnology, University of Twente, P.~O.~box 217, 7500 AE Enschede, The Netherlands}

\author{J.~F.~F.~Bulmer}
\altaffiliation{}
\affiliation{Quantum Engineering Technology Labs, University of Bristol, Bristol, United Kingdom}
\author{J.~Lugani}
\affiliation{Center for Sensors, Instrumentation and Cyber Physical System Engineering, IIT Delhi, New Delhi 110 016, India}
\author{I.~A.~Walmsley}
\affiliation{Department of Physics, Imperial College London, Prince Consort Rd., London SW7 2AZ, United Kingdom}
\affiliation{Clarendon Laboratory, University of Oxford, Parks Road, Oxford OX1 3PU, United Kingdom}

\author{P.~W.~H.~Pinkse}
\affiliation{MESA+ Institute for Nanotechnology, University of Twente, P.~O.~box 217, 7500 AE Enschede, The Netherlands}

\author{J.~Eisert}
\email{jense@zedat.fu-berlin.de}
\affiliation{Dahlem Center for Complex Quantum Systems, Freie Universit{\"a}t Berlin, 14195 Berlin, Germany}
\affiliation{Helmholtz-Zentrum Berlin f{\"u}r Materialien und Energie, 14109 Berlin, Germany}
\affiliation{Fraunhofer Heinrich Hertz Institute, 10587 Berlin, Germany}

\author{N.~Walk}
\email{nathanwalk.gmail.com}
%\altaffiliation{Present address: Toshiba Europe Ltd, Cambridge, UK.}
\affiliation{Dahlem Center for Complex Quantum Systems, Freie Universit{\"a}t Berlin, 14195 Berlin, Germany}

\author{J.~J.~Renema}
\email{jelmer.renema@gmail.com}
\affiliation{MESA+ Institute for Nanotechnology, University of Twente, P.~O.~box 217, 7500 AE Enschede, The Netherlands}
\affiliation{QuiX Quantum B.V., Hengelosestraat 500, 7521 AN Enschede, the Netherlands}

\date{
	\today
}

\begin{abstract}
One of the core questions of quantum physics is how to reconcile the unitary evolution of quantum states, which is information-preserving and time-reversible, with evolution following the second law of thermodynamics, which, in general, is neither. The resolution to this paradox is to recognize that global unitary evolution of a multi-partite quantum state causes the state of local subsystems to evolve towards maximum-entropy states. In this work, we experimentally demonstrate this effect in linear quantum optics by simultaneously showing the convergence of local quantum states to a generalized Gibbs ensemble constituting a maximum-entropy state under precisely controlled conditions, while introducing an efficient certification method to demonstrate that the state retains global purity. Our quantum states are manipulated by a programmable integrated quantum photonic processor, which simulates arbitrary non-interacting Hamiltonians, demonstrating the universality of this phenomenon.  Our results show the potential of photonic devices for quantum simulations involving non-Gaussian states.
\end{abstract}

\maketitle

\section*{Introduction}
\label{sec:intro}

One of the long-standing puzzles of theoretical physics is how notions of statistical physics and of basic quantum mechanics fit together in closed systems \cite{vonNeumann29}. Statistical mechanics is concerned with probabilistic, stationary ensembles that maximize entropy under external constraints. Elementary quantum mechanics, in contrast, describes the  deterministic evolution of quantum states of closed systems under a specified Hamiltonian. It has become clear \cite{1408.5148,ngupta_Silva_Vengalattore_2011,christian_review, Popescu06} that these seemingly contradictory premises can be resolved by making the distinction between \textit{global} unitary dynamics and \textit{local} relaxation (see Fig.~1). The physical mechanism is that local expectation values converge to those of statistical ensembles, while the entire closed quantum system undergoes unitary dynamics. Large-scale, closed quantum systems therefore appear locally thermal without the need to postulate an external heat bath. Crucially, this local equilibration behaviour is believed to be ubiquitous, in the sense that 
one has to fine-tune the Hamiltonian in order to not observe it \cite{Popescu06,Linden:2009ii}.

The mechanism of local equilibration is particularly clear-cut under non-interacting quadratic bosonic Hamiltonians, such as describe linear quantum optics. If the initial state is non-Gaussian, it is expected to `Gaussify' in time, i.e., to locally converge to Gaussian states that maximize the entropy given all second moments of the state \cite{AnalyticalQuench,CramerCLT,GluzaEisertFarrelly,SchmiedmayerGaussian,PhysRevE.100.022105}. 
In this case, for local Hamiltonian dynamics, it can be rigorously proven \cite{AnalyticalQuench,CramerCLT,GluzaEisertFarrelly,PhysRevE.100.022105}
that the state converges to a so-called 
\emph{generalized Gibbs ensemble} (GGE) \cite{Rigol_etal08,GeneralizedGibbs,AnalyticalQuench,Wouters2014a,CalabreseEsslerFagotti11}, i.e., a thermal ensemble under further constants of motion or conserved charges.
Notwithstanding this comparably clear theoretical situation, only very recently, there has been substantial experimental progress \cite{langen_2015,kaufman_2016,SchmiedmayerGaussian,LongExperimentalGaussification,GGEWork3}, with still not all aspects being clarified.
This is primarily due to the fact that it is challenging to create sufficiently isolated experimental systems to rule out that the observed equilibration is not due to decoherence, but in fact to the desired dynamics \cite{langen_2015,neill_2016, Trotzky,gring_2012,monroe_2016,clos_2016,Islam:2015cm}.
	
In this work, we experimentally show universal, reversible
equilibration and Gaussification, using an integrated quantum photonic processor (see Fig.~\ref{fig:fig2_setup}), i.e., a programmable linear optical interferometer. We use the very high degree of control available in integrated photonics to simulate for arbitrary interaction times a large number of randomly chosen quadratic Hamiltonians, including ones that are not restricted to nearest-neighbour coupling. We exploit the size of the optical network to implement a set of additional optical transformations that certify that the observed relaxation is due to the internal dynamics of our multi-mode quantum state and not due to interaction with the environment, by undoing the Hamiltonian. We find that the single-mode measurements converge to those of a thermal state with a temperature corresponding to the mean photon number, while the overall time evolution can be undone, which certifies universal, reversible Gaussification. These results exemplify the advantages of photonics as a platform for quantum simulation  \cite{PhotonicQuantumSimulators, BristolQuantumSimulation,Peruzzo,PanSampling, Szameit, PhysRevA.97.062304,PhysRevLett.120.130501}, namely good scaling of decoherence with system size, a high degree of experimental control, and the rapid growth in achievable quantum systems, both measured in the number of optical modes and in the number of photons. The fact that photonic quantum interference without explicit photon-photon interactions carries computational hardness, as demonstrated by the hardness of boson sampling \cite{aaronson_computational_2010,lund_boson_2014,hamilton_gaussian_2017} shows that even non-universal photonic processors can perform operations beyond the capabilities of classical devices \cite{PanSampling,zhong_quantum_2020,zhong_phase-programmable_2021,Dhand}. The technological contribution of this work is to go a substantial step further and investigate to what extent the newly found levels of control and system size can be exploited for photonic quantum simulation of systems of interest, contributing to placing integrated optical devices in the realm of quantum technological devices \cite{Roadmap,PhotonicQuantumSimulators,Pitsios:2017by,Barz:2015hx,Ma:2014dv} for quantum simulation.  
\smallskip

 \begin{figure}
    \centering
\includegraphics[width=.36\textwidth]{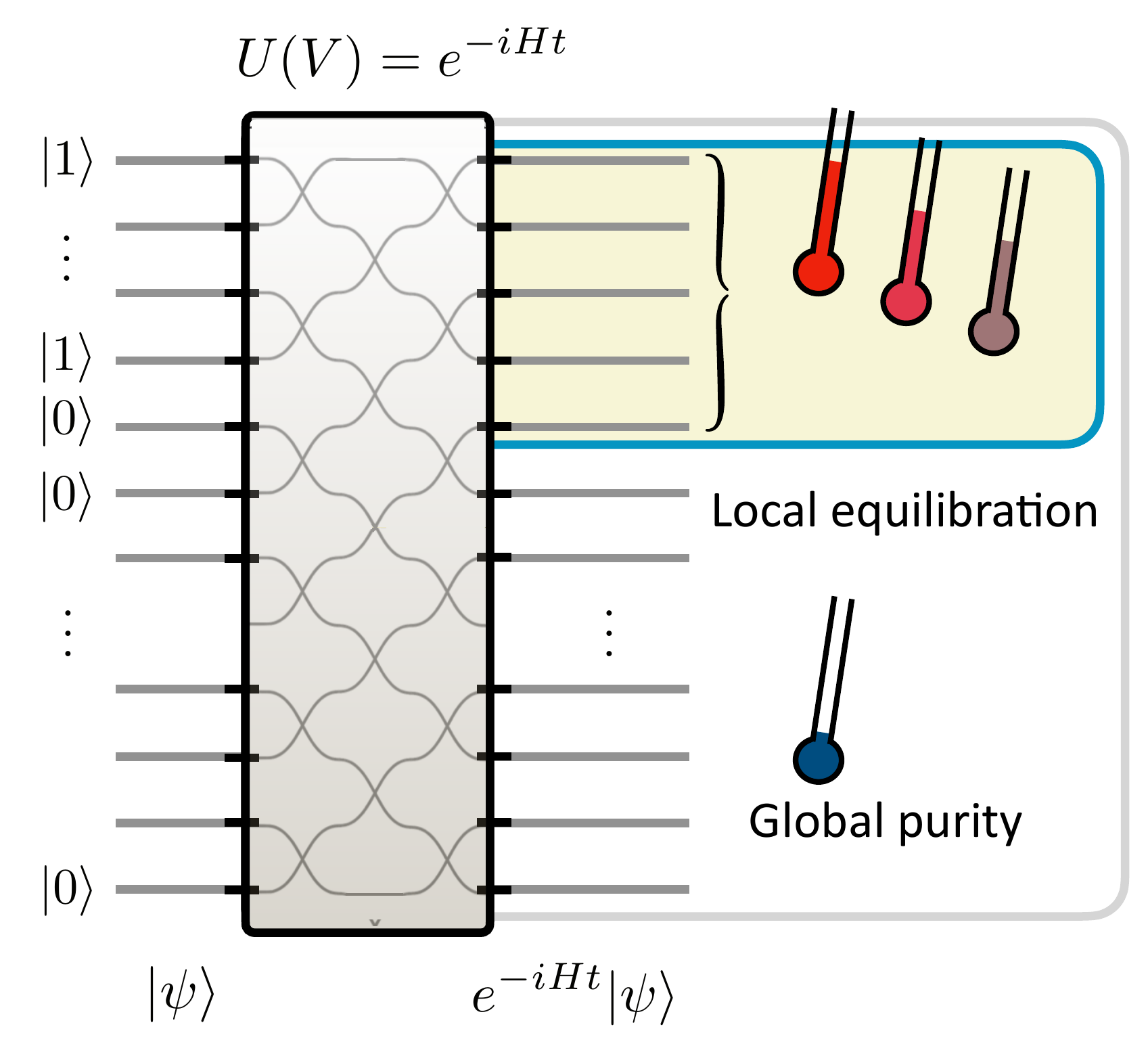}
    \caption{Photonic simulation of quantum equilibration. A closed, many-body quantum system, initialized in a product state 
    and undergoing unitary evolution generated by a Hamiltonian, necessarily remains in a pure state. However, local observables may exhibit a generalized thermalization. Entanglement builds up between sub-systems until after some time $t_{\mathrm{eq}}$, each sub-system appears to have approximately relaxed into a maximum entropy state. The paradigmatic case of a non-Gaussian bosonic state evolving under a quadratic Hamiltonian can be probed via a photonic simulation platform. A fully programmable linear optical chip can provide `snapshots' of the local and global system dynamics for arbitrary times and interaction ranges by implementing the appropriate unitary %$U = e^{iHt}$. 
    $U(V) = e^{-i\hat Ht}$ with $V\in {\rm U}(m)$ for
    $m$ modes.}

    \label{fig:fig1_schem}
\end{figure}

\section*{Results}

\subsection*{Local equilibration}
In any setting governed by closed-system Hamiltonian dynamics, equilibration can only happen locally for local observables, since the global entropy must be preserved in time. In the setting considered, the global system is a multi-mode linear-optical system initially prepared in a highly non-Gaussian state $\rho$ on $m$ bosonic degrees of freedom, namely  $|\psi\rangle\langle\psi|$ with $|\psi\rangle =|1,\dots, 1,0,\dots, 0\rangle$	of $n = 3$ single photons in $m = 4$ optical modes. 
The bosonic modes are associated with annihilation
operators $\hat{b}_1,\dots, \hat{b}_m$.
The subsequent integrated linear optical circuit is given by a unitary $V\in {\rm U}(m)$ that linearly transforms the bosonic modes. Any unitary from the
group ${\rm U}(m)$ of $m\times m$ unitary matrices
can be realized by a suitably designed 
linearly optical circuit. In state space, such linear optical circuits are reflected by $\rho\mapsto \sigma:= U(V) \rho U(V)^\dagger$, where $U(V)$ is the physical implementation of the passive  mode transformation $V$ 
that linearly transforms a set of bosonic
operators to a new set as
$(\hat{b}_1,\dots, \hat{b}_m)^T \mapsto
V (\hat{b}_1,\dots, \hat{b}_m)^T 
$. The representation of the mode transformation in Hilbert space $V\mapsto U(V)$
is commonly referred to the metaplectic representation in technical terms. 
Finally, the output distribution is measured in the Fock basis using quasi-photon-number-resolving detectors, giving measurements of the form 
$\mu\mapsto P(\mu)$ with
\begin{equation}
	    P(\mu) = \langle n_1,\dots, n_m| U(V) \rho U(V)^\dagger|
	    n_1,\dots, n_m\rangle,
\end{equation}
where $\mu=(n_1,\dots, n_m)$ 
is a given pattern of detection events.
 
For our purpose of showing local equilibration, we interpret the evolution $U(V)=e^{-i \hat Ht}$ as the evolution under a Hamiltonian $\hat H$ for time $t>0$, which distributes information. In the linear optical system at hand, we will implement two Hamiltonians, a quadratic bosonic translationally invariant `hopping' Hamiltonian, resembling the non-interacting limit of a Bose-Hubbard Hamiltonian, and a Haar random transformation $V\in {\rm U}(m)$ corresponding to a Hamiltonian with random long-range interactions. In a fixed-size optical system, we can simulate the evolution at various times by tuning the strength of the evolution, interpreting $t$ as scaling the strength rather than the duration of the interaction.

As the time $t$ gets larger, increasingly longer-ranged entanglement builds up. This means that the expected moments of the local photon number $
\hat n_j:= \hat{b}_j^\dagger \hat{b}_j$ 
of each of the output modes
labeled $j=1,\dots, m$ 
of the state $\sigma$ will increasingly, in the depth of the circuit,  
equilibrate and lead to a distribution that resembles that of 
a (generalized) Gibbs ensemble. In other words, as
seen in Fig.~1, one encounters local equilibration where the reduced quantum states
of a subset of the modes, or individual modes, 
equilibrate and take thermal-like values. Equivalently, we can say that the state will locally thermalize, in the sense that it results in the same expectation values for local observables as if the entire system had relaxed to a thermal equilibrium state.

Strictly speaking, here we observe a generalized thermalization in the following sense. The Gibbs or canonical state reflecting thermal equilibrium is given by $\xi := e^{-\beta \hat H}/{\rm tr}(e^{-\beta \hat H})$ for a
suitable inverse temperature $\beta>0$ that 
is set by the energy density. For non-interacting bosonic 
systems, local equilibration for subsystems consisting of several modes is instead expected to 
converge to a generalized Gibbs state. To be specific, here, the initial state is a product state (and hence has obviously short-ranged correlations) -- albeit not being translationally invariant -- and the bosonic quadratic
Hamiltonian will on the one hand be translationally invariant before
it undergoes a time evolution generated by
$U(V) = e^{-i \hat Ht}$ (or the  Haar-random $V\in {\rm U}(m)$). 
The situation is particularly transparent where $\hat H$ is a hopping Hamiltonian which is translationally invariant. Defining
the momentum space occupation numbers as
\begin{equation}
    \hat N_k:= \frac{1}{m}
    \sum_{x,y=1}^m 
    e^{2\pi i k (y-x)/m}
    \hat{b}_x^\dagger \hat{b}_y
\end{equation}
one finds that the generalized Gibbs ensemble
is then given by the maximum-entropy state
$\omega$ given by
 \begin{equation}
 	\omega:= \text{argmax}\{S(\eta):
 	\text{tr}(\eta \hat N_k)
 	=
 	\langle\psi| \hat N_k
 	|\psi\rangle
 	\text{ for all }k
 	\},
 \end{equation}
associated with an inverse temperature per momentum mode, where $S(\eta) = -\mathrm{tr}(\eta \log \eta)$ is the von Neumann entropy. 
For an infinite system, convergence to such a state is guaranteed \cite{AnalyticalQuench,CramerCLT, GluzaEisertFarrelly,PhysRevE.100.022105}, in the sense that the global pure state will remain pure, but
again, all reduced states (and for that matter, all expectation values of local observables) will
for most times take the values of this generalized Gibbs ensemble.
For finite systems, it has been rigorously settled in what sense the state is locally approximated by such a \emph{generalized Gibbs ensemble} \cite{Rigol_etal08,GeneralizedGibbs,Wouters2014a,CalabreseEsslerFagotti11,AnalyticalQuench} before recurrences set in. We discuss the specifics of this mechanism in more detail in Supplementary Note 2. For the Haar-random unitaries, we still find Gaussification in expectation, creating an interesting state of affairs, as here the theoretical underpinning is less clear. 

For subsystems consisting of a single bosonic mode only, canonical or Gibbs states as well as generalized Gibbs ensembles both give rise to identical photon number distributions reflecting Gaussian states: The state `Gaussifies' in time. The situation at hand is particularly simple in the situation where the expectation value of the photon number is the same for each of the $m$ output modes. Then for a Gaussian state,
 the probability of observing 
 $k$ photons reduces to
 \begin{equation}
    p(k) = \frac{\binom{n-k+m-2}{n-k}}{\binom{n+m-1}{n}} =
	 \frac{D^k}{(D+1)^{k+1}} \left\{ 1 + {O}\left(\frac{1}{m}\right)\right\},
	 \label{eqn:theoryeqn}
 \end{equation}
where $D := {n}/{m}$ is the photon density per mode, which acts as an effective temperature. 
 
Interestingly, generalized Gibbs ensembles are still not quite thermal or canonical Gibbs states, which would be maximum-entropy states given the expectation value of the energy, but a generalization of that state, due to the non-interacting nature of the Hamiltonian. For example, in full non-equilibrium dynamics under large-scale interacting Bose-Hubbard Hamiltonians (as can be probed with cold atoms in optical lattices \cite{Trotzky}) one expects an apparent relaxation to a Gibbs state. In contrast, a generalized Gibbs ensemble maximizes the von-Neumann entropy under the constraint of the energy expectation and the momentum space occupation numbers which are preserved under the non-interacting translationally invariant evolution $t\mapsto e^{-i\hat Ht}$. Therefore, one can say that each of the momentum modes is then associated with its own temperature, as sketched in Fig.~1, and the system `thermalizes' up to the constraints of the momentum space occupation numbers being preserved. 

Such generalized Gibbs ensembles are also interesting from the perspective of quantum thermodynamics \cite{GGEWork,GGEWork2,GGEWork3}.
The presence of the additional conserved charges indeed alters the thermodynamic properties and comes in as a further constraint. It is also found that the
minimum-work principle can break down in
the presence of a large number of conserved quantities \cite{GGEWork}. 
Resource theories for 
 thermodynamic exchanges of non-commuting and hence non-Abelian observables are also strongly altered for generalized Gibbs ensembles compared to their thermal counterparts \cite{GGEWork2}.

\subsection*{Certification}
In this section, we lay out the certification tools that we have developed to verify that the experiment has worked close to its anticipated functioning. Crucially, time evolution preserves the purity of a quantum system; the system only appears to be equilibrated when considering the local dynamics. Therefore, in the ideal case, it should be possible to undo the time evolution after applying $U$. This leads to the evolution $U^\dagger U = I$, meaning that a revival of the initial, non-Gaussian state is observed. In a noiseless experiment, this operation would function perfectly, and all entanglement will be formed between the photons as opposed to between the photons and the environment. This latter form of entanglement corresponds to decoherence and cannot be time-reversed by acting only on the photons. Therefore, the extent to which one observes a revival of the initial state serves as a measure of the degree of photon-photon entanglement versus the degree of decoherence. 

We further formalize this idea in the form of a 
fidelity witness \cite{Leandro,hangleiter_direct_2016} 
that certifies the fidelity $F(\sigma,\ket{\psi_t}) = |\bra{\psi_t}\sigma\ket{\psi_t}|$ 
between the experimentally prepared state 
$\sigma$ and a pure target state described by a state vector %$\ket{\psi_t} = U\ket{\psi}$, 
$\ket{\psi_t} := e^{-i\hat Ht}\ket{\psi}$. The procedure requires a well-calibrated, programmable  measurement unitary and number resolving (but not spectral-mode resolving) detectors. It consists of two settings for the measurement unitary: the inverse of the target unitary and the inverse followed by a Fourier transform
$U_{\rm F}$. The constant number of measurement settings and polynomial classical computation resources required  mean the procedure is efficiently scalable to arbitrary system sizes. Here, we consider the specific case of witnessing against the specific target state of our experiment leaving the generalization to the supplementary material.

For the first measurement, we measure the state $U^\dag \sigma U$ in the photon number basis. More specifically, we measure the fraction $p_1$ of detection events which correspond to our input state (i.e., exactly one photon in the first three input modes and no photon in the fourth mode). 
If our photodetectors would perfectly resolve the temporal and spectral degrees of freedom of the photons, this measurement in itself would be sufficient for certification  \cite{Leandro}. However, in our system, the detectors only resolve the spatial mode. Neglecting this and naively carrying out the above procedure could result in certifying a large fidelity even with photons in distinct temporal modes, i.e., distinguishable states.

To rule this out, we employ an additional measurement setting, as part of a two-step certification process: we implement $U^\dag$ followed by a Fourier transformation and  count photons. From the first setting, we upper bound the probability
$p_1$ of seeing one photon in each of the first three spatial modes and no photon in the fourth. From the second, we upper bound $p_2$, the overlap probability of $\sigma$ with the distinguishable sub-space. This is done by monitoring the fraction of observed interference patterns that would be forbidden for truly indistinguishable photons following a Fourier transform \cite{Tichy:2010gi,Tichy:2012gt,Tichy:2014ua}.

In this way, we arrive at a fidelity bound of the form 
\begin{equation}\label{eq:fidelitybound1}
F \geq p_1 - \frac{9}{4}p_2 - \delta(\epsilon)
\end{equation}
where $\epsilon>0$ is the probability that the bound is correct, and $\delta$ is the corresponding statistical penalty, which arise from the observed photon counting statistics on $p_1$ and $p_2$. This bound is derived from Chebyshev's inequality and holds with very few assumptions on the underlying distribution (for a full derivation, see the supplemental material).

If one is merely interested in establishing the presence of entanglement in the system, 
one can derive a simple entanglement witness $\mathcal{W}$ from the estimated fidelity. We use the following definition of an entanglement witness \cite{Friis:2019hg}
\begin{equation}
    \mathcal{W}:=\lambda^2_{\rm max} \mathbb{I} - \ket{\psi_t} \bra{\psi_t}
\end{equation}
where $\lambda^2_{\rm max}$ is the maximal Schmidt coefficient in the decomposition of $\ket{\psi_t}$ over a given partition, whose classical computation is not scalable, but feasible in our case. It follows then that $F>\lambda^2_{\rm max}$ is a witness of entanglement.

\subsection*{Integrated photonic platform}

\begin{figure*}
    \centering
   
    \includegraphics[width=.9\textwidth,keepaspectratio]{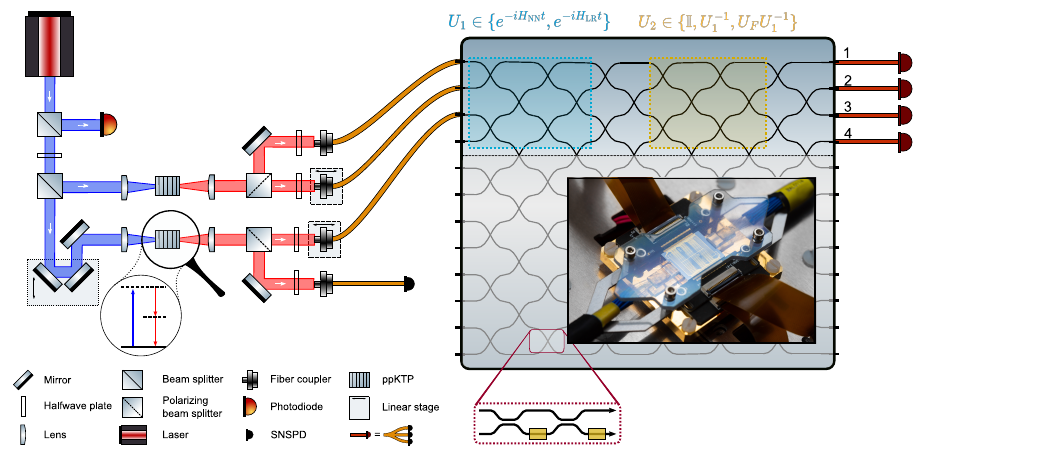}
    \caption{
    Overview of the setup.
    The 
    left hand side of the figure shows the two \emph{spontaneous parametric down-conversion} (SPDC) sources based on nonlinear \emph{periodically poled Potassium Titanyl Phosphate} (ppKTP) crystals, in which blue pump photons are spontaneously split in two red photon pairs. One of these four photons is used as a herald and the other three are injected in the first three modes of our $12\times12$ integrated
    photonic programmable processor. 
    The processor  
    output is sent to small fibre-beam-splitter networks and \emph{superconducting nanowire single photon detectors}  (SNSPDs), which act as pseudo-number counting detectors. In the processor, we program the unitary $U_1$ used to simulate the temporal dynamics (the blue block). In addition, we can program a second unitary $U_2$ for the verification process (the yellow block). The zoom-in shows a Mach-Zehnder interferometer that implements one of the programmable beam-splitters. The inset shows a photograph of a fibre-connected integrated optical chip nominally identical to the one used in the experiment. Photo credit for the inset photo: Gijs van Ouwerkerk (PHIX Photonics Assembly).} 
    \label{fig:fig2_setup}
\end{figure*}

We use an integrated quantum photonics architecture as our experimental platform (see Fig.~\ref{fig:fig2_setup}). Integrated quantum photonics constitutes a platform for non-universal quantum simulation based on bosonic interaction between indistinguishable photons  \cite{BristolQuantumSimulation,carolan2015universal,tillmann_2013,harris_2018_OpticaOPTICA,qiang_2018_Nat.Photonics,arrazola_2021_Nature}.	In integrated quantum photonics, quantum states of light are fed into a large-scale tuneable interferometer and measured by single-photon-sensitive detectors. 
	 
Our interferometer is realized in silicon nitride waveguides \cite{TriPlex, QuiX2021}, and has an overall size of $n = 12$ modes and an optical transmission of $2.2-2.7\,$dB, 
i.e., $54\%-60\%$ depending on the input channel. Reconfigurability of the interferometer is achieved by a suitable arrangement of unit cells consisting of pairwise mode interactions realized as tuneable Mach-Zehnder interferometers \cite{clements_optimal_2016}. Each unit cell of the interferometer is tuneable by the thermo-optic effect. For a full 12 mode transformation, the average amplitude fidelity $F = {n}^{-1}
\mathrm{Tr} 
(|U^{\dagger}_{\rm set}||U_{\rm get}|)$ is $F = 0.98$, where $U_{\rm set}$ and $U_{\rm get}$ are the intended and achieved unitary transformations in the processor, respectively. The processor preserves the second-order coherence of the photons \cite{QuiX2021}.
	
We implement a quantum simulation of thermalization and a verification experiment in two separate sections of the interferometer. These two sections are indicated in blue and yellow, respectively in Fig.~\ref{fig:fig2_setup}; the area below the dotted line in Fig.~\ref{fig:fig2_setup} is not used. These two sections both form individual universal interferometers on the restricted space of four optical modes, allowing us to apply two arbitrary optical transformations $U_1$ and $U_2$ in sequence. 

We use the first section to simulate time evolution of our input state. We select two families of Hamiltonians to simulate: A hopping Hamiltonian $\hat H_\text{NN}= \gamma \sum_k  \hat{b}_k^\dagger \hat{b}_{k+1} + {\rm h.c.}$ 
which consists of equal-strength nearest-neighbour interactions between all modes, which simulates the superfluid, non-interacting limit of the Bose-Hubbard model, and a set of 20 randomly chosen long-range Hamiltonians $\hat H_\text{LR}= \sum_{i,j} \gamma_{i,j} \hat{b}_j^\dagger \hat{b}_{i} + 
{\rm h.~c.}$, which we generate by applying the matrix logarithm to a set of Haar-random unitary matrices \cite{mezzadri_how_2006}.
	
The second section of the interferometer, indicated in Fig.~\ref{fig:fig2_setup} in yellow, is used for certification. When we wish to directly measure the quantum state generated by the first section, we set this area to the identity, leaving the state after $U_1$ untouched. However, we can also use this second section to make measurements in an arbitrary basis on the quantum state generated by $U_1$, which allows us to certify the closeness of our produced quantum state to the ideal case. 
	
Our photon source is a pair of \emph{periodically poled potassium titanyl phosphate} (ppKTP) crystals operated in a Type-II degenerate configuration, converting light from $775\,$nm to $1550\,$nm \cite{evans_2010_Phys.Rev.Lett.}, with an output bandwidth of  $\Delta \lambda \approx 20\,$nm. By using a single external herald detector and conditioning on the detection of three photons after the chip, we post-select on the state vector $|\psi\rangle = |1,1,1,0\rangle$ \cite{tillmann_2013}. By tuning the relative arrival times of our photons, we can continuously tune the degree of distinguishability between our photons. On-chip measurements via the \emph{Hong-Ou-Mandel} (HOM) effect \cite{hong_measurement_1987} lower bound the wave function overlap between photons $x = |\braket{\psi_i | \psi_j}|$, according to $V = x^2$, where $\ket{\psi_i}$ is the wave function of photon $i$, and $V$ is the visibility of the HOM dip. We measure visibility's of 89\% and 92\% for photons of different sources, and 94\% for photons of the same source. Photon detection is achieved with a bank of $13$ superconducting single-photon detectors \cite{ReviewSNSPD, Marsili-SNSPD}, which are read out with standard correlation electronics. For each of our four modes of interest, we multiplex three detectors to achieve quasi-photon number resolution \cite{feito_measuring_2009} with the thirteenth detector used as the herald. By means of adjusting the time delay between the photons we can adjust their degree of mutual distinguishability. We can switch between indistinguishable particles, which produce an overall entangled state (i.e., exhibiting both modal and particle entanglement), which will exhibit thermalization, and distinguishable particles, in which each photon traverses the experiment unaffected by the others, corresponding to a product state of the single-photon wave functions, which does not exhibit local thermalization.

\subsection*{Experimental results}  
\begin{figure*}
    \centering
 \includegraphics[width=1\textwidth,keepaspectratio]{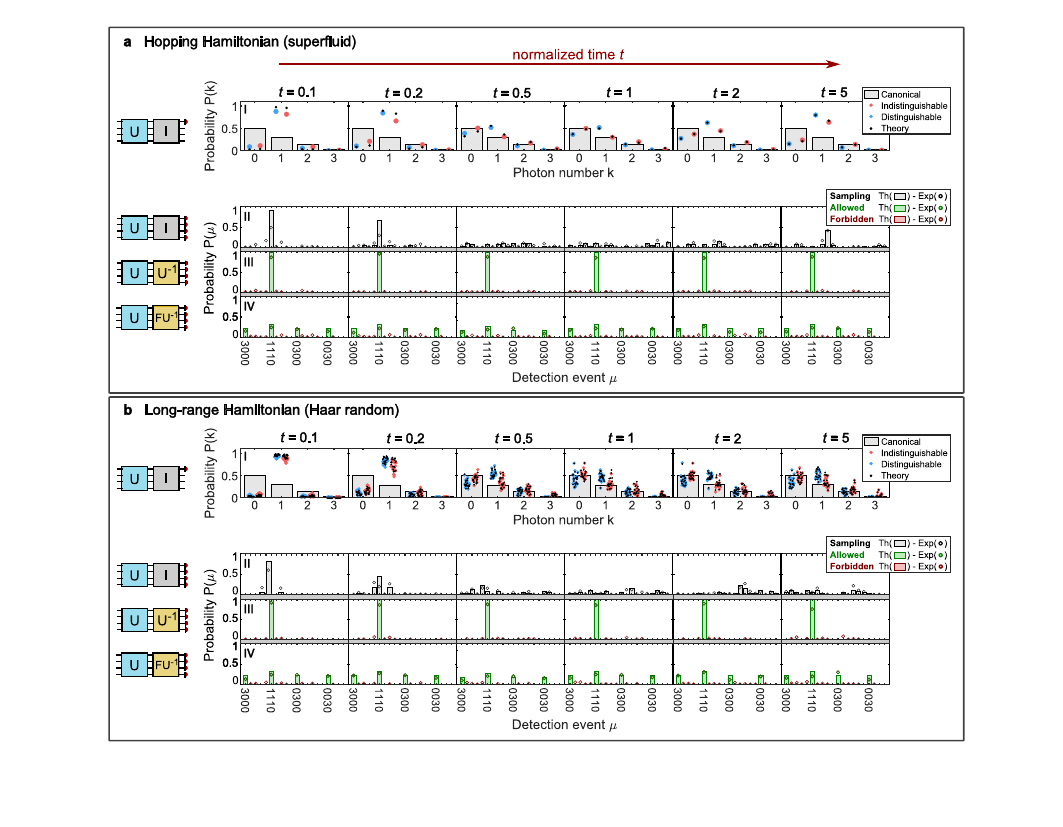}
     \caption{a) Hopping Hamiltonian: In panel I, the time evolution of photon-number probability distribution in spatial output mode $1$ is plotted. The black points (squares) show the theoretical prediction for indistinguishable (distinguishable) particles, while coloured points correspond to experimental data. Panels II-IV show the observed output distributions. These rows correspond to the output distributions of the hopping Hamiltonian (panel II), the first certification measurement $U^{-1}$ (panel III) and the second certification measurement $U_FU^{-1}$ (panel IV). Theoretical predictions (Th) are represented by bars and the experimental results (Exp) are represented by circles. The green-coloured data corresponds with outcomes that benefit the certification protocol, whereas the red data is forbidden, i.e., 
     ideally should not occur.
     b) Long-range Haar-random model: In panel I, the time evolution of photon-number probability distribution in spatial output mode $1$ for 20 different random Hamiltonians are plotted. The black points (squares) show the theoretical prediction for indistinguishable (distinguishable) particles, while coloured points correspond to experimental data. Panels II-IV show the observed output distributions for the first long-range Hamiltonian. These rows correspond to the output distributions of the first long-range Hamiltonian (panel II), the first certification measurement $U^{-1}$ (panel III) and the second certification measurement $U_FU^{-1}$ (panel IV). Theoretical predictions (Th) are represented by bars and the experimental results (Exp) are represented by circles. The green-coloured data corresponds with outcomes that benefit the certification protocol, whereas the red data is forbidden, i.e., ideally should not occur.
    } 
    \label{fig:fig3_Thermalization}
\end{figure*}

Figure~\ref{fig:fig3_Thermalization} shows the results of our quantum simulation of the hopping Hamiltonian and $20$ random instances of longe-range Hamiltonians, in sub-figures a) and b), respectively. The two sub-figures each have a tabular structure, where the columns indicate the different simulated time steps, with the simulation time indicated at the head of the column, and the rows indicate different measurement settings, i.e., either the experiment itself or the corresponding certification measurements.  The data in these figures was acquired over $20$ minutes for the photon number distribution, $320$ minutes per certification measurement for the hopping Hamiltonian and $220$ minutes for each certification measurement of the long-range Hamiltonian, with four-photon events (three photos in the processor plus herald) occurring at a rate of 4 Hz.

The first row of the two sub-figures displays the single-mode photon-number statistics $k\mapsto p(k)$ as generated after the application of $U$ in the first section of the processor. The output statistics were measured for the first output mode. The experiment was carried out for both distinguishable (blue points) and indistinguishable (red points) photons. The grey bars show the expected distribution at full equilibration given by 
Eq.~(\ref{eqn:theoryeqn}). For both Hamiltonians, initially, the input state is still clearly present, as indicated by the high probability to observe exactly one photon in the observed output mode. However, entanglement builds up as time evolves, since the photons increasingly equilibrate. Consequently, for the indistinguishable photons, the initial input state evolves to a thermal-like state at $t=1$.  For both Hamiltonians, the distinguishable photons (whose output statistics correspond to those of classical particles) do not approach the canonical thermal state, demonstrating the intrinsic link between entanglement and thermalization.

For the hopping Hamiltonian, at later times ($t = 2, t = 5$), the finite size of our Hamiltonian gives rise to a \textit{recurrence}, i.e., the state moves away from equilibrium again and evolves back towards the initial input state \cite{AnalyticalQuench,CramerCLT}. For the long-range Hamiltonian, in contrast, the long-range interactions mean that recurrences are pushed away to later time not included in the simulation. These results suggest the presence of long-range order (as opposed to structured, nearest-neighbour interactions) tends to increase the time for which a system will continue to exhibit local relaxation. Whilst this picture is intuitive, a rigorous understanding of these effects is an exciting open problem for theory and future experiments. The general agreement across a large range of randomly chosen Hamiltonians also represents strong experimental evidence for the ubiquity of these effects \cite{Popescu06,Linden:2009ii}.

The second row of the two sub-figures shows the full output-state distribution $\mu\mapsto p(\mu)$ after only the application of $U$, measured with indistinguishable photons. The bars in the background correspond to the expected distributions. For the long-range Hamiltonian, a single representative example of our 20 Hamiltonians is plotted. From this data, it can be clearly seen that at the point of thermalization, the photons are spread over many possible output configurations, whereas a recurrence manifests as a transition back to fewer possible output configurations. 

The third and fourth rows show the output-state distributions after the first and second certification measurement, respectively. In these rows, the output configurations which contribute positively to the fidelity witness are indicated in green, and those which contribute negatively are indicated in red. The first certification measurement undoes the entanglement generated by $U$ and ideally only results in
state vectors of the form $\ket{\psi_{\rm out}}=\ket{1,1,1,0}$.
The second certification measurements also applies a three-mode Fourier to the generated states. Ideally, this results in only four allowed output configurations. These certification measurements show good agreement with the ideal allowed states, demonstrating the high degree of control over the experiment. For the second certification measurement, most of the deviations from the expected distribution can be attributed to the known photon indistinguishabilities. From the data presented in the third and fourth row, we extract the values of $p_1$ and $p_2$, respectively, which are used in the fidelity witness as laid out in Eq.~(\ref{eq:fidelitybound1}). 

\begin{figure*}
    \centering
 \includegraphics[width=.8\textwidth,keepaspectratio]{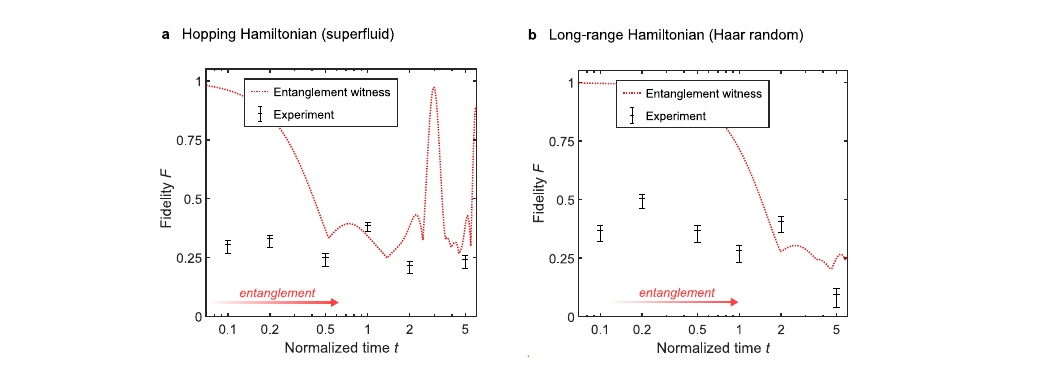}
    \caption{a) Certification of entanglement in the hopping Hamiltonian (superfluid): The lower bound certification fidelity estimations for the hopping Hamiltonian are plotted against a theoretical entanglement witness. b) Certification of entanglement in long-range Hamiltonian (Haar-random): The lower bound certification fidelity estimations for the first long-range Hamiltonian are plotted against a theoretical entanglement witness. In both plots, the top, middle and bottom points at each time step correspond to confidence values of $\epsilon = 0.7$, $\epsilon = 0.8$ and $\epsilon = 0.9$, respectively. The background colour saturation qualitatively shows the total entanglement generated at that time step, which is proportional to the value of the entanglement witness. A higher saturation indicates  a stronger presence of multi-photon entanglement. 
    } 
    \label{fig:fig4_certification}
\end{figure*}

Fig.~\ref{fig:fig4_certification}a) and \ref{fig:fig4_certification}b) show the certified fidelities for both the hopping Hamiltonian and the first random long-range Hamiltonian, respectively. The three horizontal ticks on each data point correspond to confidence values of $\epsilon = 0.7$, $\epsilon = 0.8$ and $\epsilon = 0.9$. The line shows the entanglement witness, corresponding to a bi-partition between mode 1 and the remaining modes. The relatively constant fidelity to the target global state contrasts against the conversion of the local, single-mode statistics to thermal statistics, as seen in Fig.~3.

Fig.~\ref{fig:fig4_certification}a) shows that entanglement is certified for $t = 1$ in the hopping Hamiltonian system. The observed fidelity $F=0.359$ is above the threshold of the entanglement witness. Similarly, 
Fig.~\ref{fig:fig4_certification}b) shows a unambiguous certification for the first long-range Hamiltonian at $t = 2$. The fidelity $F=0.360$ is well above the certification threshold. Both of these entanglement certifications hold with a confidence of at least 90\%.

The certification fidelities are limited by imperfect control over the processor. This follows from the certification fidelity at $t=0.2$ for the long-range Hamiltonian. This fidelity $F= 0.462$ is significantly higher than others. Closer inspection shows a near optimal value for $p_2$, which is now only limited by the partial distinguishability of the generated photons. This implies that the certification at other time steps is limited by imperfect chip control, i.e., a limited fidelity at which any measurement can be implemented. 
A second factor limiting the certification is detector blinding, which affects the obtained values of $p_1$ (see the supplemental material for more details on detector blinding and the convergence of the certification statistics).

% \{Outlook}	
\section*{Discussion}
In conclusion, we have experimentally shown that a pure quantum state in a closed environment can locally behave like a thermal state because of entanglement with the other modes. To this end, we simulated both the non-interacting limit of a Bose-Hubbard hopping Hamiltonian and $20$ random long-range Hamiltonians on a programmable $12$-mode photonic processor. Previous experiments in this direction have not been able to show this kind of reversibility since creating a sufficiently isolated quantum system and controllable evolution is notoriously difficult. However, our experiment is fully time-reversible, just like quantum mechanics itself. This reversibility has allowed us to certify that equilibration and thermalization are due to entanglement between the quantum particles rather than with the environment. These results also provide experimental evidence for the universality of these phenomena and shed new light on the role of long-range interactions on relaxation dynamics. From the point of view of the development of quantum technologies, these experiments showcase the degree of control, low decoherence and rapidly growing size of integrated quantum photonic processors as instances of a near-term quantum computational platform.

% Not finished - FS, 2022-08-23
\section*{Methods}
\subsection*{Photon source and input state preparation}
Distinguishable and indistinguishable photonic quantum state vectors of the form $\ket{\psi} = \ket{1,1,1,0}$ are generated by a multi-photon source consisting of two free-space Type-II SPDC sources. Two non-linear $2\,$mm length ppKTP crystals (Raicol Crystals) are pumped by a Ti:Sa mode-locked laser (Tsunami, Spectra Physics) at $775\,$nm with a spectral bandwidth of $5.4\,$nm FWHM. Pulses are generated with a repetition frequency of $80\,$MHz and $150\,$fs pulse duration. Each crystal is pumped by approximately $10\,$mW pump power, generating degenerate signal-idler pairs at $1550\,$nm with generation probability $<1\%$ per pulse. Typical heralding efficiencies for individual crystals are around $40 – 45$\%, while typical two-photon event rates are $\sim 0.20$ MHz coincidence counts at $40\,$mW pump power. While the source is designed to produce as pure photons as possible, residual energy and momentum conservation result in spectral signal-idler correlations. These correlations are attributable to the periodically-poled structure of the non-linear crystals. We suppress these correlations by using a spectral bandpass filter of $\Delta \lambda = 25\,$nm. Halfwave plates are used to remove the distinguishability in photon polarization and to match the TE mode supported by our quantum photonic processor. Three motorized linear stages (SLC-$2475$, Smaract GmbH) are used to control relative photon arrival times, used to switch the distinguishability of the photons.

\subsection*{Quantum photonic processor}
Our quantum photonic processor consists of a photonic chip, the control electronics which actuate this chip, and peripheral systems such as cooling. The photonic chip implements arbitrary linear optical transformations on 12 waveguides. The waveguides are implemented as stoichiometric silicon nitride ($\mathrm{Si_3N_4}$) \emph{asymmetric double-stripe} (ADS) waveguides with the TriPleX technology \cite{TriPlex}. The waveguides are optimized for light of a wavelength of 1550 nm, and have propagation loss of $< 0.1$ dB/cm. The waveguides have a minimal bending radius of 100 um. Coupling on and off the chip is achieved by adiabatic mode converters, which are implemented by removing the top layers of the ADS stack. These converters have coupling losses down to 0.9 dB / facet. The overall measured loss budget of the processor is $2.5 \pm 0.2$ dB, with roughly 1.8 dB attributable to the two adiabatic couplers and 0.7 dB to propagation losses on chip.%, and the remainder to the bulkhead connectors at the front panel of the processor box.\\ 
\\
\\
Universality of the optical transformation is achieved by a network of beam splitters in a checkerboard geometry. Each tuneable beam splitter is implemented as a \emph{Mach-Zehnder interferometer} (MZI), with two static 50/50 directional couplers. To tune the MZI, two thermo-optical phase shifters are used, one inside the MZI which enables shifting of light amplitude between adjacent optical modes, and one external to the MZI which allows for a relative phase shift between the two modes. The thermo-optic phase shifters are implemented as 1 mm long platinum heaters, and have $V_{\pi} = 10$ V, and dissipate roughly 400 mW of power each. This power is carried off the chip through a Peltier element which is itself actively cooled with water cooling. A bank of 132 digital-to-analog converters converts signals from a control computer to voltages over the heaters. A dedicated software package is used for communication, and to compute the required voltages. Control over the processor to the precision required in this experiment requires understanding of the crosstalk between these control channels, which is achieved in a dedicated software package.

\subsection*{Photon detection system}
A suite of $13$ superconducting nanowire single-photon detectors (SNSPDs) is used for photo-detection. These detectors are biased close to their critical current ($8$ to $22$ $\mu$A range), operating at quantum efficiencies of around $90\%$ for $1550\,$nm photons with typical $200\,$Hz dark counts. Fourfold coincidence rates within a $750\,$ps window are monitored by a time tagger device (Timetagger Ultra, Swabian). From the combination of photon generation rates and dark count rates, we estimate that less than one in a million measured four-fold coincidence events are expected to be triggered by a dark count. Polarization maintaining (PM) fibres are used in combination with polarization controllers to optimize and stabilize output counts in each channel. Pseudo-number resolution detection is realized by multiplexing detectors in a 
\emph{1-to-3 quasi-photon number resolving detector} (q3PNRD) configuration by fibre beam splitters on the four optical modes of interest, with the thirteenth detector used as a herald. 

\subsection*{Photon detection calibration}
In order to  sample from $\mu\mapsto P(\mu)$ in an unbiased way, as required in this work, it is important to characterize the relative output losses from the different detectors. The SNSPDs have variation in their detection efficiency, and the same holds for the output coupling of the various optical modes of the photonic processor. Non-uniformity in the overall detection efficiency of our experiment biases the sampling of $P(\mu)$, since it will suppress some outcomes while relatively enhancing others. Note that this does not hold for any inhomogeneities in the in-coupling, due to post-selection. Furthermore, we assume that on-chip losses are reasonably uniform, which is evidenced by the high matrix amplitude fidelities. Furthermore, note that an absolute detection calibration (a notoriously difficult problem at the single-photon level) is not necessary, only a relative one between the $12$ detectors of interest.

Non-uniform detection channel losses are characterized by directly transmitting heralded single photons from input mode 1 to all four output modes consecutively; these optical transformations can be performed with high fidelity. In each of these four consecutive experiments, the heralded singles count rate of each detector in the q3PNRD behind the output mode of interest is measured. All measured heralded singles count rates $S_{i}$ originate from the same on-chip uniform heralded single photon rate $R_{1}$, therefore, it is convenient to pool all other losses such as out-coupling efficiencies, detection efficiencies and splitting ratios for each detection channel $i$ in a lumped factor $p_i$, to get
\begin{equation}
    S_{i} = p_{i}R_{1}.
\end{equation}
Since we are only interested in relative efficiencies, we introduce relative weight factors for each detection channel, which are then normalized with respect to the maximum measured heralded singles rate and defined by
\begin{equation}
    w_{i} = \frac{S_{i}}{S_\text{max}}.
\end{equation}
In our experiments, we achieved excellent weight factor stability. Typically, we observed less than 1\% relative fluctuations over more than 15h time span.
\smallskip

Similar to nonuniform detection efficiency, the fact that each qPNRD is effectively less efficient when detecting multiple photons as opposed to a single photon biases the output distribution and must be corrected for. 
Experimentally, we measure heralded threefold coincidence rates $CC_{p,q,r}$, which denote the rate at which detectors $p$, $q$ and $r$ and the herald detector fire simultaneously, normalized to the overall frequency of successful experiments. The challenge is then to convert these probabilities into an unbiased estimate of $P(\mu)$. 

To compensate for q-PNR effects, we enumerate all combinations of threefold detection event which would give rise to a particular output pattern $\mu$. For probabilistic multi-photon detection, the probability of measuring $j$ photons behind mode $i$ when $k$ photons are injected is denoted $P_i(j|k)$. We note that for $P_i(1|1)$ and $P_i(2|2)$ there are three possible permutations, while for $P_i(3|3)$ there is just one permutation. More explicitly, we find
\begin{equation}
    P_i(0|0) = 1,
\end{equation}
\begin{equation}
    P_i(1|1) = w_{p_i} + w_{q_i} + w_{r_i},
\end{equation}
\begin{equation}
    P_i(2|2) = 2!(w_{p_i}w_{q_i} + w_{q_i}w_{r_i} + w_{p_i}w_{r_i}),
\end{equation}
\begin{equation}
    P_i(3|3) = 3!w_{p_i}w_{q_i}w_{r_i},
\end{equation}
where $w$ are the weight factors determined above and $w_{p_i} + w_{q_i} + w_{r_i} \leq 1$ due to incorporated losses. Since all $P_i(j|k)$ are independent probability events, we find for an estimate for $P(\mu)$
\begin{equation}
    P(\mu) = \frac{ \sum_{(p,q,r)\in \mu} CC_{p,q,r}  }{P_1(n_1|n_1)P_2(n_2|n_2)P_3(n_3|n_3)P_4(n_4|n_4)},
\end{equation}
where $\mu=(p,q,r) $ denotes all combinations of detection events contributing to the same $\mu$ and $n_i$ is the number of photons detected in a mode $i$ for a given $\mu$. These results are used to correct raw measurement data.
\smallskip

\subsection*{Data availability} 
All experimental and simulated data used in this study are available in the 4TU.ResearchData database \cite{Somhorst2023}. 

\subsection*{Code availability} 
All data post-processing and simulation code used in this study are available in the 4TU.ResearchData database \cite{Somhorst2023}.

\begin{widetext}
\newpage

\section*{Supplementary Note 1 -- Derivation of fidelity witness}
A fidelity witness provides guarantee that the fidelity of some target state with an experimental output is at least a certain threshold value with at least a certain probability \cite{Leandro}. Here,  we present a derivation of such a witness, including finite-size statistics, that is efficient in terms of experimental effort and classical computation. In the first place we can use this threshold as evidence that our global system retains approximately the same fidelity with a target pure state whilst the local systems exhibit apparent entropy increase. A natural further question that arises is, is a particularly meaningful fidelity threshold? In this experiment, where the key feature of interest is the role of entanglement in producing local entropy production, we will use previously established relationships between fidelity and entanglement (see, e.g., Supplementary Ref.~\cite{Friis:2019hg} to establish useful benchmarks).
The idea is that the fidelity between a separable state and an entangled target state vector $\ket{\psi_t}$ cannot exceed a certain threshold, which is set by the largest Schmidt coefficient. If the fidelity exceeds that threshold, i.e., if 
$F > \lambda^2_{\rm max}$, entanglement must be present. Because the size of the largest Schmidt coefficient decreases with the amount of entanglement, for more entangled states lower fidelities are sufficient to witness the presence of entanglement.\\

\textbf{Ideal case: Fully-mode-resolving detectors.}
The fidelity between a quantum state $\sigma$ and a target state $\sigma_t=\ket{\psi_t}\bra{\psi_t}$ is defined as
\begin{equation}
    F(\sigma, \sigma_t) := \mathrm{tr}(\sigma\sigma_t) = \bra{\psi_t}\sigma \ket{\psi_t}.
\end{equation}
In our case, the target state vector is an initial state vector
\begin{eqnarray}
\ket{\psi} = \ket{\underbrace{1,\dots ,1}_n,\underbrace{0,\dots,0}_m}
\end{eqnarray}
of single photons in the first $n$ modes of an $m$ mode system, evolved by a unitary, $U(V)$, implementing a passive linear optical transformation $V$, i.e., $\sigma_t = \ket{\psi_t} \bra{\psi_t} = U(V)\ket{\psi)} \bra{\psi}U(V)^{\dagger}$. 
Thus, the fidelity can 
then be written as
\begin{equation}
    F = F(\sigma, U\ket{\psi} \bra{\psi}U^{\dagger}) = F(U^{\dagger}\sigma U, \ket{\psi} \bra{\psi})
\end{equation}
where we have suppressed the argument $V$ for brevity, and is lower bounded in terms of photon number operators by \cite{Leandro}
\begin{equation}
    F^{(n)} = \bigg\langle (n+1-\hat{n}) \prod_{j=1}^n \hat{n}_j \bigg \rangle_{U^{\dagger} \sigma U},
\end{equation}
where $\hat{n}_j = \sum_{n_j=0}^\infty n_j \ket{n_j} \bra{n_j}= \hat{b}_j^\dagger \hat{b}_j$ 
are the photon number operators, whose eigenvalues are the number of photons in mode $j$ and $\langle \sum_j\hat{n}_j \rangle = n$ is the global photon number.
When one post-selects for a constant global photon number in each run (which we do in our experimental setup, to $n=3$), then the bound simplifies to
\begin{equation}\label{Fn}
    F^{(n)} = \bigg\langle\prod_{j=1}^3 \hat{n}_j\bigg \rangle_{U^{\dagger} \sigma U} .
\end{equation}
 Note that the only calculation necessary is to compute the Hermitian conjugate of the given matrix $U$.\\
\\    

\textbf{Real case: Spatial-mode-resolving detectors.}
In our experimental setup we do not have fully-mode-resolving detectors, meaning that we have no physical equivalent of $\hat{n}_j$.
%= \sum_{n_j=0}^\infty n_j \ket{n_j} \bra{n_j}$. 
% ALREADY DEFINED.
%
In particular, our detectors can only resolve spatial modes and no temporal ones. This leaves an uncertainty regarding the exact mode of the photon after the measurement. Instead of projecting onto a unique mode (a single pure quantum state $M_k = \ket{k}\bra{k}$), our detectors project onto a set of states, which are spread out over the temporal degrees of freedom and, without further work at least, cannot be distinguished. This uncertainty severely limits the fidelity that can be established. For three temporal modes, the measurements at each of the four spatial modes correspond to the following set of operators,
\begin{equation}
    \begin{split}
        M_0 &= \ket{0,0,0}\bra{0,0,0},\\
        M_1 &= \ket{1,0,0}\bra{1,0,0} + \ket{0,1,0}\bra{0,1,0} + \ket{0,0,1}\bra{0,0,1},\\
        M_2 &= \ket{2,0,0}\bra{2,0,0} + \ket{0,2,0}\bra{0,2,0} + \ket{0,0,2}\bra{0,0,2}\\ &+ \ket{1,0,1}\bra{1,0,1} + \ket{1,1,0}\bra{1,1,0} + \ket{0,1,1}\bra{0,1,1},\\
        M_3 &= \ket{3,0,0}\bra{3,0,0} + \ket{0,3,0}\bra{0,3,0} + \dots,
    \end{split}\label{Ms}
\end{equation}
where $M_0$ detects the absence of 
photons and $M_1$, $M_2$ and $M_3$ measure one, two and three photons, respectively, in a given spatial mode. Since we set up our experiment such that the ideal initial and the target state consists of one photon per spatial mode, our measurement operator of interest will be $M_1$ and our certification scheme will be based on the measurement of $M_1^A \otimes M_1^B \otimes M_1^C \otimes M_0^D$, where $A$, $B$, $C$ and $D$ label the four spatial modes. We will want to verify whether the initial state is being recovered after implementing the unitary and its inverse, i.e., that the three photons are in the same temporal mode and each in a different one of the first three spatial modes. The problem with using these measurements and naively applying the bound in Supplementary Eq.~(\ref{Fn}) is that, considering Supplementary Eq.~(\ref{Ms}), it is clear these measurements can produce the ideal click pattern even if the photons were completely distinguishable and no quantum interference or multi-photon entanglement has been present.

At this point we introduce some alternative notation that will come in handy later: The numbers in the ket-vector label the occupied temporal mode (there are three temporal modes, so the numbers go from $1$ to $3$) and the subscript 1 indicates that each spatial mode is occupied by one photon only (which is the case for $M_1$). 
For example, all photons being in the first temporal mode reads $\ket{1,1,1}_1 = \ket{1,0,0;1,0,0;1,0,0;0,0,0}$. The first photon in the first, second photon in the second, and third photon in the third temporal modes reads $\ket{1,2,3}_1 = \ket{1,0,0;0,1,0;0,0,1;0,0,0}$.
    
Now that we have pointed out the ambiguity of just counting photons in spatial modes, and equipped with useful notation, we next turn to the question of how to overcome the uncertainty regarding the temporal or spectral degree of freedom. An answer lies in the observation that certain interference patterns can be clearly associated with non-synchronous, i.e., distinguishable, photon states (similar to a HOM dip \cite{hong_measurement_1987}) 
-- we call those interference patterns \textit{forbidden 
patterns}. We make use of this effect in practice by implementing a Fourier transform $U_F$ after the unitary $U$ and its inverse and use the overlap between the distinguishable subspace of states and the image of the forbidden patterns under the Fourier transform $U_F$ to sharpen the lower bound on the fidelity. It is worth mentioning that this overlap is not 1:1 and some ambiguity will remain. It does, however, reduce the ambiguity significantly and thereby increases the estimated fidelity in a useful way.
    
We now give a full derivation of the fidelity bound. The bound, as in Main Eq.~(5), has two components. 1) A lower bound $p_1$ on seeing one photon per spatial mode and 2) an upper bound $p_2$ on the overlap of $\sigma$ with the distinguishable subspace. %Together these two pieces of information give insights into i) whether the implemented unitaries are correct and ii) the indistinguishability of the photons . 
The two components correspond to two different measurement settings:
\begin{enumerate}
    \item Implement the unitary and its inverse and count photons.
    \item Implement the unitary and its inverse, implement a Fourier 
    transform (an interference 
    experiment) and then count photons.\\
\end{enumerate}

\textbf{First measurement setup.}
For simplicity and without loss of generality we fix $M_1^A$ to the first temporal mode, i.e., $M_1^A=\ket{1,0,0}\bra{1,0,0}$. The first measurement setup can be expressed as the operator product $\ket{1,0,0}\bra{1,0,0} \otimes M_1^B \otimes M_1^C \otimes M_0^D$. The overlap of the state $\sigma$ with $\ket{1,0,0}\bra{1,0,0}\otimes M_1^B \otimes M_1^C \otimes M_0^D$, i.e., the probability of seeing one photon in each of the first three spatial modes (regardless of the temporal modes), can be estimated experimentally with accuracy $\epsilon_1$. This estimation takes the form of a lower bound $p_1$ -- the result of the first round of measurements.
\begin{equation}\label{eq:boundone}
    \text{tr}[U^\dag\sigma U (\ket{1,0,0}\bra{1,0,0}\otimes M_1^B \otimes M_1^C \otimes M_0^D)] \geq p_1
\end{equation}
which expanded yields
\begin{equation}\label{eq:fidbound}
    \begin{split}
        &\text{tr}\big[U^\dag \sigma U( \\
        &\ket{1,1,1}_1\bra{1,1,1}_1+ \color{blue}\ket{1,1,2}_1\bra{1,1,2}_1 + \ket{1,1,3}_1\bra{1,1,3}_1+\\
        &\color{blue}\ket{1,2,1}_1\bra{1,2,1}_1 + \ket{1,3,1}_1\bra{1,3,1}_1 + \color{blue}\ket{1,2,2}_1\bra{1,2,2}_1 +\\  
        & \color{blue}\ket{1,3,3}_1\bra{1,3,3}_1\color{red}+\ket{1,2,3}_1\bra{1,2,3}_1+\ket{1,3,2}_1\bra{1,3,2}_1 \color{black})\big] \\
        &\geq p_1.
    \end{split}
\end{equation}
The first term (black) reflects the fidelity given by $F=\text{tr}\big[U^\dag \sigma U(\ket{1,1,1}_1\bra{1,1,1}_1)\big]$. The other terms correspond to the overlap of $U^\dag \sigma U$ with those states where one (blue) or two (red) photons are distinguishable. We summarize those states as $\color{blue}\hat{P}_1$ and $\color{red}\hat{P}_2$, respectively, and call $\hat{P}=\color{blue}\hat{P}_1+\color{red}\hat{P}_2$ the distinguishable subspace. With that shorthand notation Supplementary 
Eq.~(\ref{eq:fidbound}), 
simplifies to
\begin{equation}\label{eq:fidbound_short}
    F \geq p_1 - \text{tr}\big(U^\dag \sigma U \hat{P}\big).
\end{equation}
We can calculate $p_1$ by counting the relative number of instances of 
$M_1$ in our first measurement setup.\\
\\

\textbf{Second measurement setup.}
Next, in order to improve the fidelity bound in Supplementary Eq.~(\ref{eq:fidbound_short}), we need to upper bound its second term $\mathrm{tr}(U^\dag \sigma U \hat{P})$, which enters the fidelity bound with a minus sign. This term contains the overlap of $\sigma$ with the distinguishable subspace $\hat{P}$. We upper bound it by implementing a Fourier interference experiment on the first three modes. This is described by the unitary $U_F = U^{\mathrm{Four}}_3\otimes 
\mathbb{I}$, where $U^{\mathrm{Four}}_n(V^{\mathrm{Four}}_n)$ %\je{[why $n$?]}
is the appropriate Hilbert space operator corresponding to the physical implementation of the mode transformation
\begin{equation} \label{eq:Fourjk}
    (V^{\mathrm{Four}}_n)_{j,k} = \frac{1}{\sqrt{n}}e^{i {2\pi}(j-1)(k-1)/n}.
\end{equation}\\
Ideally, the first three modes should each be occupied by one perfectly indistinguishable photon (the first term in Supplementary Eq.~(\ref{eq:fidbound})). For a state of this form, some counting patterns corresponding to projections onto certain output states are impossible  \cite{Tichy:2010gi,Tichy:2012gt,Tichy:2014ua}. By contrast, as we will now show, situations involving anything other than the ideal case \emph{will} result in some forbidden measurement outcomes with a certain probability 
(Supplementary Fig.~\ref{fig:fig5_forbidden}). 
Working backwards from this, we can use the observed frequency of the forbidden states to get a worst case upper bound on $\mathrm{tr}(U^\dag \sigma U \hat{P})$. This kind of discrimination is something we could not do using only the first setup. In this case, the forbidden patterns of photon numbers are all those which correspond to state vectors not included in the set $\{|1,1,1,0 \rangle, |3,0,0,0 \rangle, |0,3,0,0 \rangle, |0,0,3,0 \rangle\}$ \cite{Tichy:2010gi,Tichy:2012gt,Tichy:2014ua}. Defining the projection onto all forbidden states as $M_f$, the quantity obtained with our second measurement setup can be written as
\begin{equation}
     \mathrm{tr}\big(U_FU^\dag \sigma U U_F^{-1} M_f \big) = \mathrm{tr}\big(U^\dag \sigma U U_F^{-1} M_f U_F \big) \leq p_2,
\end{equation}
where $p_2$ is the the probability of observing a forbidden state. Using the cyclicity of the trace in the middle term we see we can also interpret $p_2$ as the overlap of $U^\dag \sigma U$ with the image of the forbidden states based on the photon counting measurements. 
    
\begin{figure*}
    \centering
    \includegraphics[width=.5\textwidth,keepaspectratio]{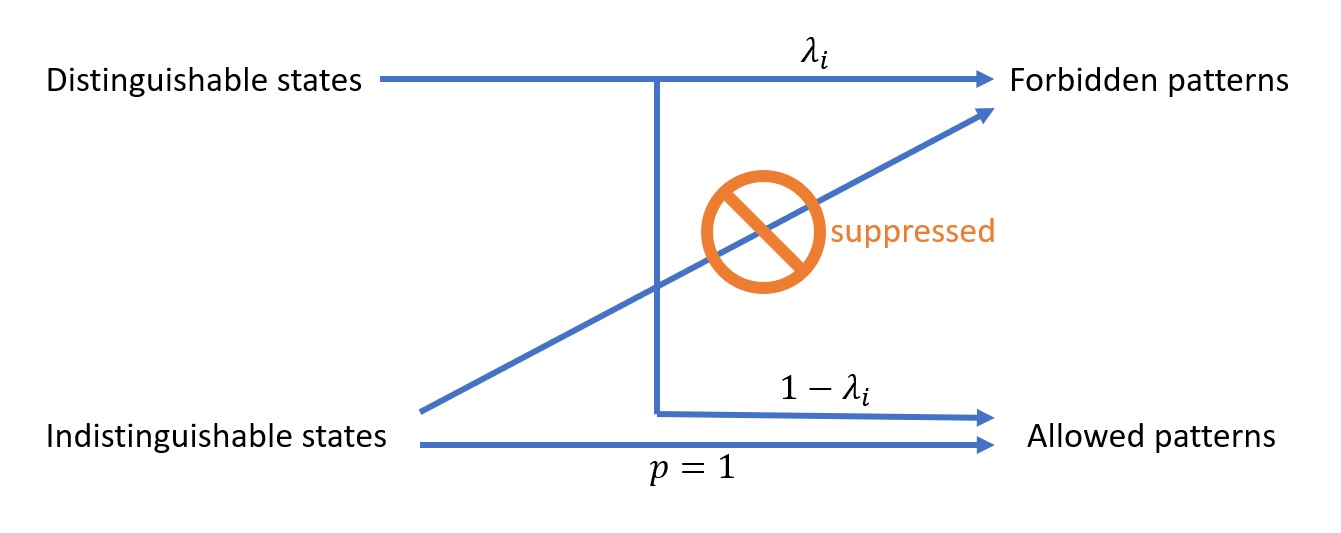}
    \caption{\textbf{Forbidden measurement probability.} This figure shows how distinguishable and indistinguishable states translate to forbidden and allowed interference patterns after the Fourier transform. Forbidden patterns are strictly suppressed for indistinguishable states (in fact they are asserted via a suppression rule), hence indistinguishable states result in allowed patterns with unit probability. Distinguishable states result in forbidden patterns with probability $\lambda_i$, which we compute for all n-photon distinguishable states in all distinguishable sub-spaces $P_i$ to establish a bound on $\text{tr}(U^\dag \sigma U \hat{P})$, the overlap between the state $U^\dag \sigma U$ and the distinguishable subspace $\hat{P}$.} 
    \label{fig:fig5_forbidden}
\end{figure*}

As mentioned before, some states within that image clearly correspond to distinguishable states, that is, states in $\hat{P}$ and other do not. To be more clear, we can express the image operator in terms of the distinguishable sub-spaces $P_i \in \hat{P}$ ($i\in \{1,2\} $) as
\begin{equation}
    U_F^{-1} M_f U_F = \sum_i\lambda_iP_i + \hat{P}_\perp
\end{equation}
where $\lambda_i :=  {\rm tr}(P_i U_F^{-1} M_f U_F)$ is the probability that a state in the distinguishable subspace $P_i \in \hat{P}$ results in a forbidden pattern and $\hat{P}_\perp$ projects onto the complementary space to $\hat{P}$. Since $\mathrm{tr}(U^\dag \sigma U \hat{P}_\perp)\geq 0$, we have
\begin{equation}\label{eq:overlapbound3}
      \mathrm{tr}\big(U^\dag \sigma U \sum_i\lambda_iP_i\big)  \leq
     {\rm tr}\big(U^\dag \sigma U U_F^{-1} M_f U_F \big) \leq p_2.
\end{equation}

To calculate these probabilities what really matters is the number of mutually distinguishable photons. For example, to calculate the probability of the forbidden state vector
$\ket{2,0,1,0}$ from states in ${\color{blue}P_1}$ with one photon distinguishable from two other identical photons), one first computes the probability of the indistinguishable photons transforming through the Fourier transform
into a state that could be transformed into a forbidden pattern by the third photon. This can be computed via Supplementary Eq.~(\ref{eq:indperm}) derived in Supplementary Ref.~\cite{Tichy:2012gt}. In this case, that would be permutations of 
the outputs $\ket{1,0,1,0}$ or $\ket{2,0,0,0}$. Then one computes the probability that the final, distinguishable photon, would fall in the correct mode to produce the forbidden output. For example, if the indistinguishable photons evolve the state
vector $\ket{1,0,1,0}$, then the distinguishable boson evolving to either $\ket{1,0,0,0}$ or $\ket{0,0,1,0}$ would result in a forbidden output pattern. The transition probabilities for distinguishable and indistinguishable photons through a Fourier transform are highly symmetric. In the above example, it turns out to makes no difference in which mode the one distinguishable boson initially resides. Working through each of the states in ${\color{blue}P_1}$ and ${\color{red}P_2}$ we find they all result in the same forbidden state probabilities of $\lambda_1 = \frac{4}{9}$ and $\lambda_2 = \frac{2}{3}$ respectively. Substituting into Supplementary Eq.~(\ref{eq:overlapbound3}) we have
\begin{eqnarray}
 \mathrm{tr}\big(U^\dag \sigma U \big(\frac{4}{9}P_1 + \frac{2}{3}P_2\big)\big) &\leq& p_2, \\ 
 \Rightarrow \mathrm{tr}\big(U^\dag \sigma U \big(P_1 + \frac{3}{2}P_2\big)\big) &\leq& \frac{9}{4}p_2 ,
 \label{eq:overlapbound2}
\end{eqnarray}
where we have divided through by $\frac{4}{9}$. We can use Supplementary Eq.~(\ref{eq:overlapbound2})
as an upper bound on $\text{tr}\big(U^\dag \sigma U \hat{P}\big) =  {\rm tr}\big(U^\dag \sigma U (P_1 + P_2)\big)$, which is what we originally set out to do. We find that
\begin{equation}
     {\rm tr}\big(U^\dag \sigma U \hat{P}\big) \leq  {\rm tr}\big(U^\dag \sigma U (P_1 + \frac{3}{2}P_2)\big) \leq \frac{9}{4}p_2
\end{equation}
and thus arrive at the following final expression for the fidelity bound in Supplementary Eq.~(\ref{eq:fidbound_short}), to get
\begin{equation}
    F \geq p_1 - \text{tr}\big(U^\dag \sigma U\hat{P}\big) \geq p_1 -\frac{9}{4}p_2.
\end{equation}
Note that this step renders the lower bound loose in general, and will tend to provide a pessimistic estimate of the fidelity.
Finally, we turn to the question of finite-size statistics.  Many tools have been developed for this situation, and here we will make use of a result from Supplementary Ref.~\cite{Leandro} based on Chebyshev's inequality which states that given $k$ independent samples and an observed fraction $p_1$ we can say that the `true' probability of that outcome $\bar{p}_1$ must satisfy
\begin{equation} \label{Chebyshev}
   \mathrm{Pr} \left[ |\bar{p}_1 -p_1 | \leq \delta \right ] \geq \varepsilon, \hspace{2mm} \delta(\varepsilon)^2 = {\frac{2
   \Sigma}{k \log(1/\varepsilon)}},
\end{equation}
where $0 \leq \Sigma <\infty$ is the variance of the distribution and $k$ is the number of measurements. This can be put together to obtain Main Eq.~(5) which holds with probability  $\epsilon=\epsilon_1 \epsilon_2$, and $\delta(\epsilon) = \delta(\epsilon_1) + \delta(\epsilon_2)$ arising from applying Supplementary Eq.~(\ref{Chebyshev}) to the experimental observations 
of $p_1$ and $p_2$.\\

\textbf{Generalization to larger systems.} This certification method can be extended to arbitrarily many modes and photons. Whilst a detailed investigation of the robustness and performance of this method is beyond the scope of this work, we briefly explain how the protocol generalizes and make some comments. The scheme can be used to certify the creation the fidelity multi-partite entangled
state vectors 
$\ket{\psi_t} = U_\mathrm{LO}(V) \ket{\psi}$ created by acting an $m$-mode linear optical unitary on an initial state vector $\ket{\psi}$, 
where the first $n\leq m$ modes are populated with indistinguishable photons as there exist forbidden states for arbitrary $n$. In fact, the technique can be slightly generalized further to any initial state where the photons are arranged in a periodic pattern.
The generalised expression for the first measurement setting would read,
\begin{eqnarray}
\mathrm{tr} \left [ U^\dag \sigma U \left (\ket{\Psi}\bra{\Psi} + \sum_i \hat{P}_i \right )  \right ] \geq p_1
\end{eqnarray}
leading to a bound
\begin{eqnarray}
F \geq p_1 - \mathrm{tr}\left (U^\dag \sigma U \sum_i \hat{P}_i \right ),
\end{eqnarray}
where the $P_i$ are all the different sub-spaces corresponding to the existence of different numbers bosons partitioned into different distinguishable `species'. There can be up to $n$ of species (i.e., one distinguishable, two distinguishable, $\dots$ , $n$ distinguishable -- if the number of species is equal to the number of photons $n$, then all photons are mutually distinguishable). 

The second measurement setting is already described for arbitrary $P_i$ and hence $n$ in Supplementary Eq.~(\ref{eq:overlapbound3})
and, recalling that 
\begin{equation}
\mathrm{tr}\left (U^\dag \sigma U \sum_i \hat{P}_i \right ) \leq \mathrm{tr}\left (U^\dag \sigma U \sum_i \lambda_i \hat{P}_i \right ), \forall \lambda_i\geq1 
\end{equation}
allows us to obtain the bound
\begin{eqnarray}
F \geq p_1 - \frac{p_2}{\min_i \lambda_i} \label{eq:fgeneral}
\end{eqnarray}
in the general case. This scheme is manifestly efficient in the number of measurement settings (two) and also scales well in terms of the the total sample size for each probability estimate. However, to evaluate the bound we naturally need to know the value of the $\lambda_i$ and also the set of forbidden states to determine $p_2$. These calculations are a one-off cost in the sense that it need only be performed once ahead of time for any value of $n$ and can then be used to certify all states in the corresponding class. In this sense, it is not counted in the scaling cost of the protocol, nevertheless it is a non-trivial overhead and we discuss the calculation in some more detail.

To explain things further we briefly recall some notation and results from
Supplementary Refs.~\cite{Tichy:2010gi,Tichy:2012gt,Tichy:2014ua}. Let $r=(r_1, r_2, \dots, r_m)$ and $s=(s_1, s_2, \dots, s_m)$ be the input and output mode occupation list with $\sum_i s_i = \sum_i r_i = n$. In our case, we have $m=4$ modes and $n=3$ photons. Our input mode occupation has been $r=(1,1,1,0)$. A useful alternative notation for the mode occupation list is the \textit{mode assignment list} $d(q)$, which is structured in terms of photons rather than modes. Its entries represent the photons and the numerical value indicates the mode that is being occupied by that photon (the list has as many entries as there are photons as opposed to as many entries as there are modes). For example, the mode occupation list $r=(2,0,0,1)$ becomes $d(r)=(1,1,4)$ (the first and second photon being in mode one and the third photon in mode four). The general expression for the mode assignment list given a mode occupation list $q$ reads 
\begin{equation}
    d(q) = \bigoplus_{j=1}^m \bigoplus_{k=1}^{q_j} (j) = (\underbrace{1,\dots,1}_{q_1},\underbrace{2,\dots,2}_{q_2},\dots,\underbrace{m,\dots,m}_{q_m}).
\end{equation}

For bosons, the transition probabilities through a Fourier transform are proportional to the permanent of an $n \times n$ sub-matrix $M$ of the $m \times m$ Fourier matrix $V^{\mathrm{Four}}_n$. With this notation of mode assignment lists, 
we can neatly express the transition probabilities as
\begin{equation}\label{eq:indperm}
    P(r,s,V^{\mathrm{Four}}_n)= \frac{|\text{perm}(M)|^2}{\prod_j r_j! s_j!}
\end{equation}
for the case of indistinguishable photons and 
\begin{equation}
    P(r,s,V^{\mathrm{Four}}_n)= \frac{\text{perm}(|M|^2)}{\prod_{j=1}^n s_j!}
\end{equation}
for distinguishable ones. The $n \times n$ matrix constructed from $V^{\mathrm{Four}}_n$, referred to as $M$, is defined as
\begin{equation}
    M_{j,k}:=(V^{\mathrm{Four}}_n)_{d_j(r),d_k(s)}
\end{equation}
where $d_j(r)$ is the $j^{\mathrm{th}}$ element of the mode assignment list $d(r)$ and the elements of $V^{\mathrm{Four}}_n$ are given in Supplementary Eq.~(\ref{eq:Fourjk}).

The forbidden patterns can efficiently be calculated as the strictly suppressed output states of the indistinguishable case with respect to the chosen Fourier transform $V^{\mathrm{Four}}_n$. More precisely, in Supplementary 
Ref.~\cite{Tichy:2010gi} it has been shown that, for a given (potentially 
$p$-periodic) initial state $r$, final states $s$ are suppressed 
through quantum interference when the criterion 
\begin{equation}
    \text{mod}\bigg(p \sum_{j=1}^N d_j(s) , n \bigg) \neq 0
\end{equation}
holds, i.e., if the above criterion holds, then the transition probability $P(r,s,V^{\mathrm{Four}}_n)$ in Supplementary Eq.~(\ref{eq:indperm}) vanishes. Having found the suppressed (i.e., forbidden) patterns, one can then go ahead and compute the $\lambda_i$ (probability that states in the various $P_i$ would result in a forbidden state). As explained above, for a given $P_i$ one needs to consider the probabilities that the populations of the  distinguishable species can combine to result in a forbidden state. Strictly speaking, to evaluate Supplementary Eq.~(\ref{eq:fgeneral}) we only need the value of the smallest $\lambda_i$. Based on preliminary investigations we conjecture that the case of $n-1$ indistinguishable bosons and 1 distinguishable boson is the minimal case. If it were necessary to check all of the $\lambda_i$, it is not trivial to determine how many calculations this would entail as it corresponds to the problem of placing $n$ indistinguishable objects in $k$ indistinguishable boxes (the bosons in each species are of course mutually distinguishable, here we are using indistinguishable in the sense that the situation that the arrangement with, say, 3 bosons in species 1 and 2 bosons in species 2 is, for our purposes, equivalent to 3 bosons in species 2 and 2 bosons in species 1) for which there is no compact form. A crude upper bound for a given $n$ and an number of species would be to count the number of ways of placing $n$ objects in $k$ \emph{distinguishable} boxes which would upper bound the number of $\lambda_i$ to be calculated via $\sum_{j=2}^n \binom{n-1}{j-1}$. Even if our conjecture is true, calculating a single $\lambda_i$ would still involve evaluating the transition probability (and hence matrix permanent) for $n-1$ bosons, which is classically hard in general. Nevertheless, to our knowledge the hardness for the specific case of a Fourier transform remains open, which leaves the total complexity of this calculation unclear for the present.

%% SUPPLEMENTARY NOTE 2
\newpage
\section*{Supplementary Note 2 -- Gaussification}
For completeness, in this subsection we recall some previous work on Gaussification and present some numerical results illustrating approximate Gaussification for the systems considered in this work. For non-interacting quadratic bosonic Hamiltonians, such as describe linear quantum optics experiments of the kind considered here, the mechanisms for equilibration have been well studied \cite{AnalyticalQuench,CramerCLT,GluzaEisertFarrelly,SchmiedmayerGaussian,PhysRevE.100.022105}. In particular, it has been rigorously shown that systems will tend to `Gaussify', meaning that after a sufficiently long enough time has elapsed, any subsystem (or even a block of subsystems) will converge to Gaussian, maximum entropy states and remain there. For finite number of modes and bosons we will never find the subsystems in perfectly Gaussian state, nor will they remain in such a state indefinitely. Instead the system will approximately Gaussify with the closeness of the approximation depending upon the system size. 

More formally, consider a state vector $\ket{\psi} \in \mathcal{H}_U = \mathcal{H}_S\otimes\mathcal{H}_E$ of the entire `universe' of our experiment which comprises $m$ modes/sites which we can think of as a system $S$ and environment $E$ with given second moments in the creation annihilation operators $\langle \hat{b}_i^\dagger \hat{b}_j\rangle>$ (i.e., the photon occupancy). Define a reduced state of a
subsystem $\varrho_S =  \ptr{\ket{\psi}\bra{\psi}}{E}$. In our work, we have been focusing on the state of this  subsystem and considering just a single mode, but one could also think of a larger subsystem. We are then interested in the dynamics as a function of time and system size. Taking the thermodynamic limit will involve fixing ratio of photons to modes ($n/m$) and then considering the limit $m\rightarrow \infty$. In that case we  want to know if $\varrho_S$ will eventually equilibrate to  $\varrho_S^{\mathrm{mc}}$, the micro-canonical state on the subsystem given the constraints on the second moments. In this limit, for fixed second moments, the maximum entropy state is a Gaussian state $\varrho_S^{\mathrm{me}} = \varrho_G$. Approximate Gaussification can then be expressed as the condition that, for any $S$ and any $\varepsilon>0$ there exists an $m$ and relaxation and recurrence times $t_{\mathrm{Rec}}$ and $t_{\text {Relax }}$, such that
\begin{eqnarray}
\left\|\varrho_{S}(t)-\varrho_{G}\right\|_{\mathrm{tr}}<\varepsilon \quad \text { for } t \in\left[t_{\text {Relax }}, t_{\mathrm{Rec}}\right]. \label{approxGauss}
\end{eqnarray}

Even for the modest system sizes at play in this work, it is still possible to observe substantial Gaussification. In Supplementary Fig.~\ref{fig:Gaussification} we numerically calculate the Wigner functions for each of the 4 modes for the initial input states at $t=0$ and an evolved state at $t=1$ for both the hopping Hamiltonian and one of the Haar-random, long-range Hamiltonians. Whilst the initial state exhibits substantial non-Gaussianity and Wigner negativity for the modes initially occupied with a single photon, after evolution all modes appear as approximately Gaussian with additional modulation caused by finite-size effects and all Wigner negativity has vanished. It is interesting to note that mode 4, which has initially been in a perfectly Gaussian vacuum state, is technically less Gaussian after evolution. Nevertheless, the system still exemplifies the phenomena described in Supplementary Eq.~(\ref{approxGauss}), as Gaussification is a 
claim that must hold for \emph{all} subsystems and not just one. After evolution the condition that all modes can be well-approximated by a Gaussian is satisfied, whereas it is radically violated by the initial input state.

\begin{figure*}[h]
\begin{center}
\includegraphics[width=0.75\textwidth,keepaspectratio]{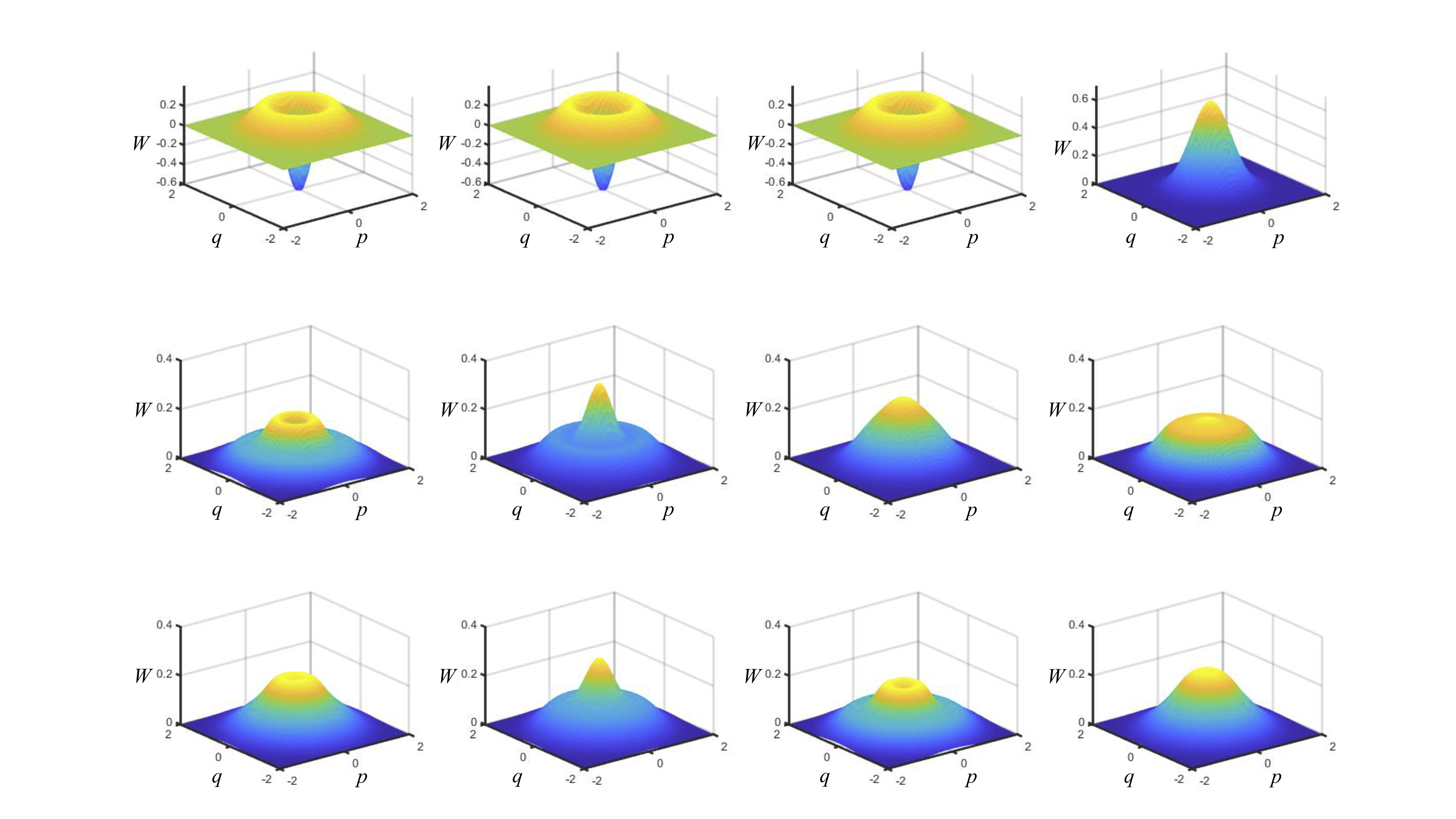}  
    \caption{\textbf{Approximate Gaussification.} Theoretical Wigner functions of each mode (columns) for the initial input state at $t=0$ (top row) and the evolved state ($t=1$) for the hopping Hamiltonian (middle row) and a long range Hamiltonian (bottom row). The Wigner quasiprobability \textit{W} is plotted as a function of dimensionless continuous momentum \textit{p} and position \textit{q} eigenvalues.}
\label{fig:Gaussification}
\end{center}
\end{figure*}

\newpage
% SUPPLEMENTARY NOTE 3
\section*{Supplementary Note 3 -- Indistinguishable photon quality}
To characterize the quality of our prepared indistinguishable $\ket{1,1,1,0}$ input state vector, we have measured  HOM-dips (see Supplementary Fig.~\ref{fig:HOMdips}), using our processor as a beam splitter between pairs of modes. We find HOM visibilities of $V_{s_1,i_2} = |\braket{\psi_ {s_1}  | \psi_{i_2} }|^2 = 89.1\%$, $V_{s_1,s_2} =|\braket{\psi_ {s_1}  | \psi_{s_2} }|^2 = 92.3\%$, and $V_{s_2,i_2} = |\braket{\psi_ {s_2}  | \psi_{i_2} }|^2 = 94.3\%$.
\begin{figure}[h]
\begin{center}
    \includegraphics[width=0.6\textwidth]{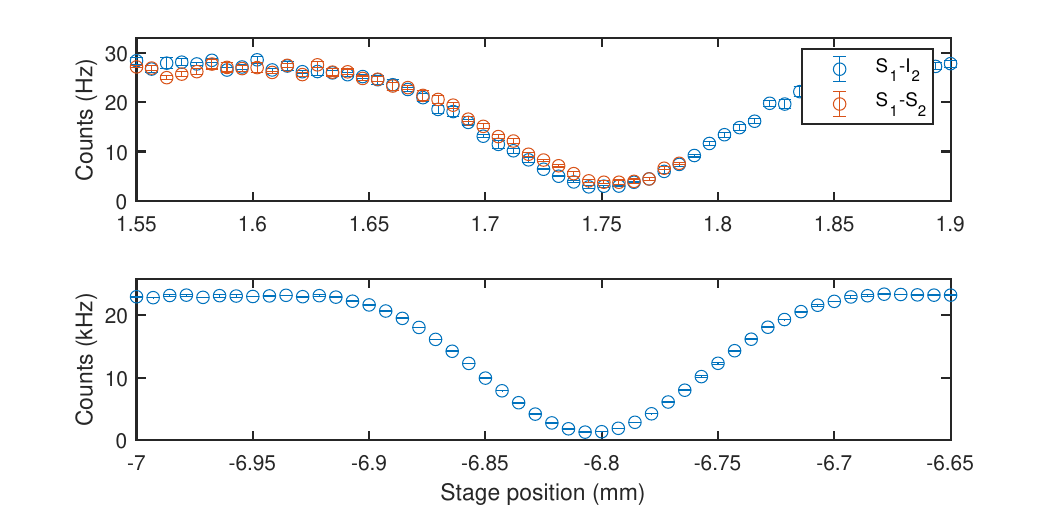}
    \caption{\textbf{Source pre-characterization.} Top panel: typical intermodal HOM dips result in visibilities of $0.89$ (signal crystal 1 $s_1$ - idler crystal 2 $i_2$) and $0.92$ (signal crystal 1 $s_1$ - signal crystal 2 $s_2$) for photons generated in different crystals. Bottom panel: the HOM visibility of a pair of photons generated by crystal $2$ (signal crystal 2 $s_2$ - idler crystal 2 $i_2$) is typically around $0.94$. Error bar represents the standard error based on 5 measurements.}
\label{fig:HOMdips}
\end{center}
\end{figure}

\section*{Supplementary Note 4 -- Quantum photonic processor operation fidelity}
To test the fidelity of the implemented optical transformations, we perform a calibration experiment. For this, classical CW light from a 1550 nm super luminescent diode (Thorlabs S5FC1005P) is injected into the input modes and an array of calibrated photodiodes (Thorlabs FGa01FC) are used to detect the output signal. The amplitude fidelities are defined as $F:=\frac{1}{n} \mathrm{Tr}(|U_{\rm set}^{\dagger} ||U_{\rm get} |)$, where $U_{\rm get}$ denotes observed transfer matrix, $U_{\rm set}$ is the target transfer matrix and the absolute signs indicate the element-wise absolute value of the matrix elements, and $n = 12$ modes is the size of the transfer matrices. For a set of 150 random permutation matrices, a value of $F = 0.992 \pm 0.002$ is found, whereas for a set of 100 Haar-random matrices we find $F = 0.979 \pm 0.01$. The full histograms of these measurements are shown in 
Supplementary Fig.~\ref{fig:Supmatfid}.

\begin{figure}[h]
\begin{center}
    \includegraphics[width=0.26\textwidth]{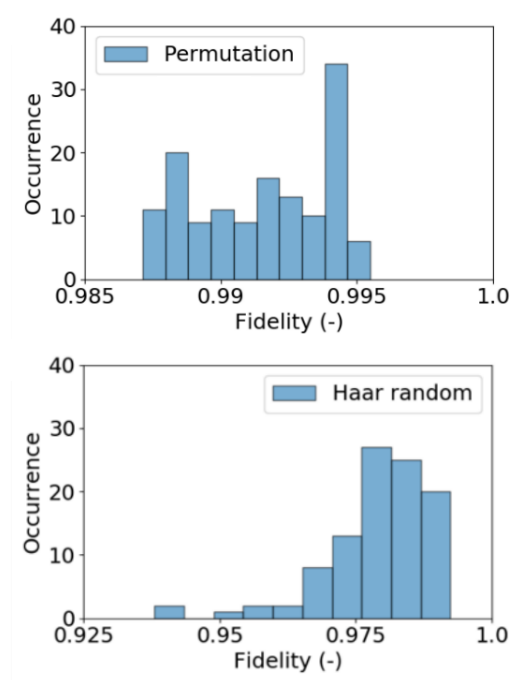}
    \caption{\textbf{Matrix fidelities.} Histogram of amplitude fidelities over two families of random matrices. Top panel: fidelity of random permutation matrices. Bottom panel: fidelity of random matrices.}
\label{fig:Supmatfid}
\end{center}
\end{figure}

\newpage
%SUPPLEMENTARY NOTE 5
\section*{Supplementary Note 5 -- Photon detector blinding}
Although post-selection on heralded three-photon events allows for extracting events based on the input state vectors $|\psi\rangle = \ket{1,1,1,0}$, other, unwanted states are frequently produced because of the probabilistic nature of SPDC sources, 
e.g., when one source produces a photon 
pair but the other one does not.
Unwanted by-product states cause detector blinding, which in combination with imperfect matrix fidelities biases observed photon statistics. This effect is illustrated in Supplementary Fig.~\ref{fig:PumpNoise} for the identity matrix transformation. Therefore, the observed photon statistics is dependent on used pump power levels. Especially photon statistics for unitary matrix transformations close to the identity matrix transformation are affected as can be seen in Main 
Fig.~3, as detector blinding is more likely due to most of the light being directed to a limited set of detectors.

\begin{figure}[h]
    \centering
    \includegraphics[width=0.5\textwidth,keepaspectratio]{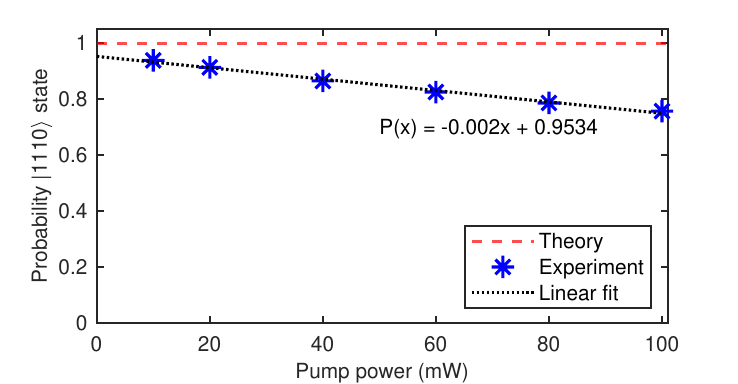}
    \caption{\textbf{Detector blinding as function of pump power.} The horizontal axis shows the pump power, the vertical axis shows the probability of observing the outcome $\mu = (1,1,1,0)$ when the identity matrix is dialled on chip, which is the only expected outcome in this case. For higher pump-power levels, the detrimental effect of by-products is increased. In case of perfect realized unitary fidelity, measurement probability is independent of pump power. Empirically, we find this effect is well described by $P(\mu) = -0.0020\cdot P_\text{\rm pump} + 0.9534$ 
    for  $\mu= (1,1,1,0)$
    with pump power $P_\text{pump}$ in mW. Error (standard deviation based on Poisson statistics) is within symbol size.}
    \label{fig:PumpNoise}
\end{figure}

\begin{figure*}[h]
    \centering
\includegraphics[width=.8\textwidth,keepaspectratio]{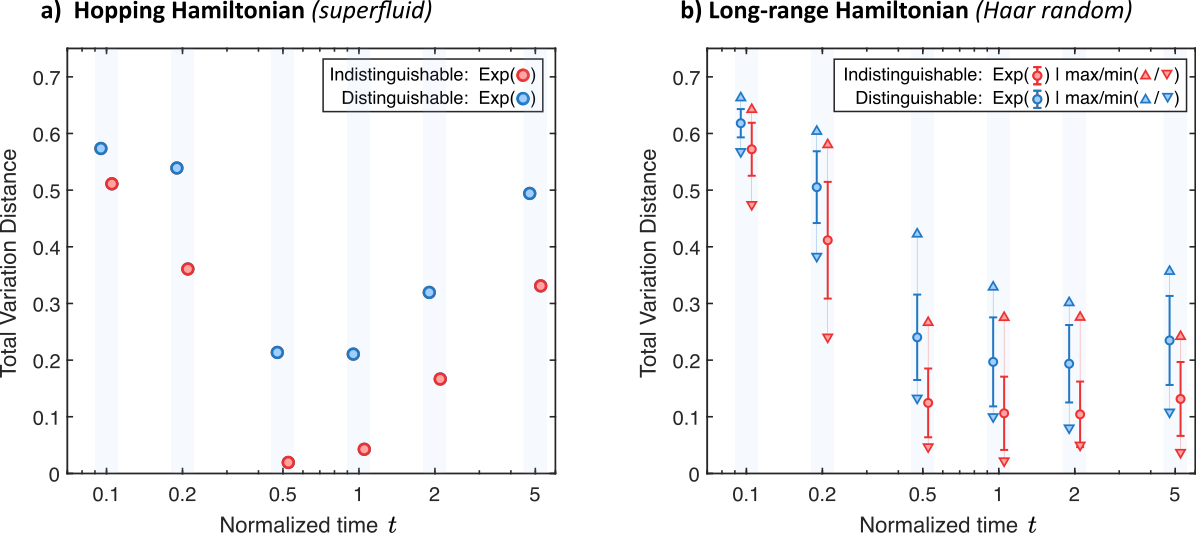}
    \caption{\textbf{Total variational distance (TVD) to the canonical thermal state as a function of normalized time $t$.}
    The red and blue symbols correspond to indistinguishable and distinguishable photons, respectively. The left panel 
    (a) shows the TVD for the six experimentally (Exp) simulated time steps on the hopping Hamiltonian. The right panel 
    (b) shows the TVD for the six time steps experimentally (Exp) simulated for the set of long-range Hamiltonians. In this panel, the average, minimum (min), maximum (max) TVD, as well as the standard deviation (error bar) in the TVD are shown.} 
    \label{fig:fig6_totalVarDist}
\end{figure*}

\newpage
\section*{Supplementary Note 6 -- Detailed experimental results}
In this section, we provide additional information on the results of our experiments. To better illustrate the convergence to Main Eq.~(4), the total variation distance between the experimental data and the canonical probability density function (from the first panel of Main Fig.~3a and 3b) is shown in Supplementary Fig.~\ref{fig:fig6_totalVarDist}. Furthermore, we consider the achievable fidelity bound as a function of measurement time. The certification fidelity only becomes meaningful when the probability of error $1-\epsilon$ is sufficiently small. This requires prolonged measurement times to accumulate sufficient statistics. Moreover, detector blinding (see Supplementary 
Fig.~\ref{fig:PumpNoise}) limits the average pump power for the fidelity certification measurements to only $5\,$mW per crystal. Such low pump power results in a fourfold coincidences rate of around $4\,$Hz. To prevent a bias caused by long term drift, all certified time steps are measured 'interleaved', i.e., each certification measurement is repeated throughout multiple times for short run time.

Supplementary Figs.~\ref{fig:BH_certification_linearized} and \ref{fig:Haar_certification_linearized} depict how the certification fidelities converge when the total measurement time is increased. Each data point is the result of a $20\,$minute measurement per certification step, i.e., $40\,$minutes per data point. The horizontal axis is linearized as $1/\sqrt{T}$, where $T$ is the measurement time in hours. Empirically, we find that the certified fidelity decreases linearly on this scale (i.e., increases linearly with $-1/\sqrt{T}$). This is consistent with the independent nature of the separate experimental runs. The red horizontal dashed lines are the bi-partition (spatial mode 1 vs spatial modes 2,3,4) fidelities required to certify entanglement. Finally, the blue solid line is a linear fit through the data points. The fit is extrapolated to $100\,$hours of measurement time. The number at the end (left) of the fit is the corresponding maximum fidelity expected based on this extrapolation.

\begin{figure}[h]
    \centering
    \includegraphics[width=.6\textwidth,keepaspectratio]{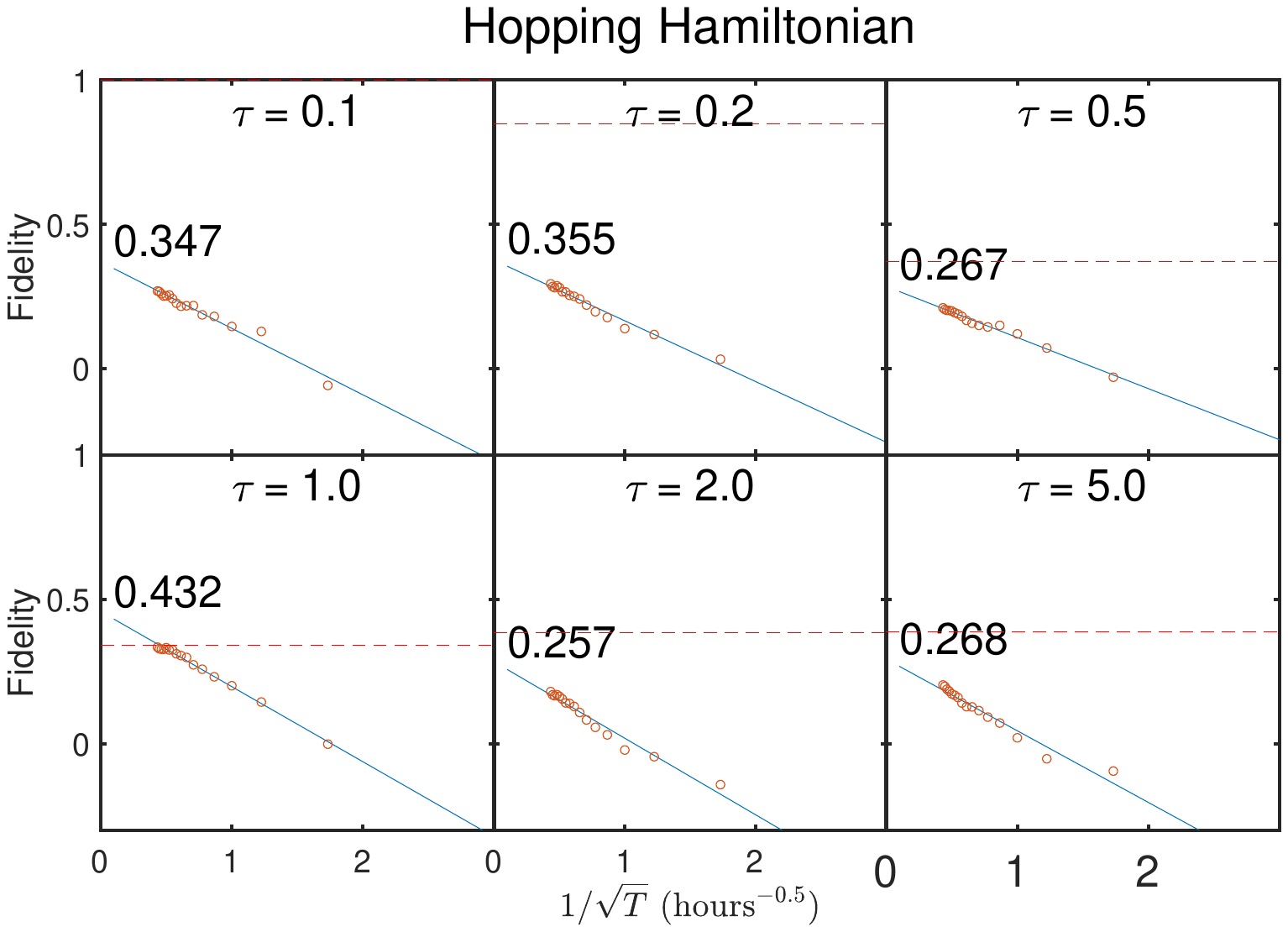}
    \caption{\textbf{Convergence of certification statistics}. Superfluid/short-range order system certification as a function of measurement time for all 6 normalized time steps $\tau$, for confidence level $\epsilon = 0.9$. $T$ indicates the measurement time.}
    \label{fig:BH_certification_linearized}
\end{figure}

Supplementary Fig.~\ref{fig:BH_certification_linearized} shows the convergence of the certification fidelity for the non-interacting 
Bose-Hubbard, or superfluid, Hamiltonian. Each panel corresponds with one of the six simulated time steps. There are a total of $16$ batches for each time step, which is almost sufficient to certify $\tau=1.0$ against the bi-partition with $\epsilon = 0.9$. Therefore, we included $10$ more batches measured under similar conditions to increase the fidelity for $\tau=1.0$ from $F = 0.335$ to $F = 0.359$. Furthermore, the extrapolations indicate that longer measurement times are not going to certify the remaining simulated time steps.

\begin{figure}[h]
    \centering
    \includegraphics[width=.6\textwidth,keepaspectratio]{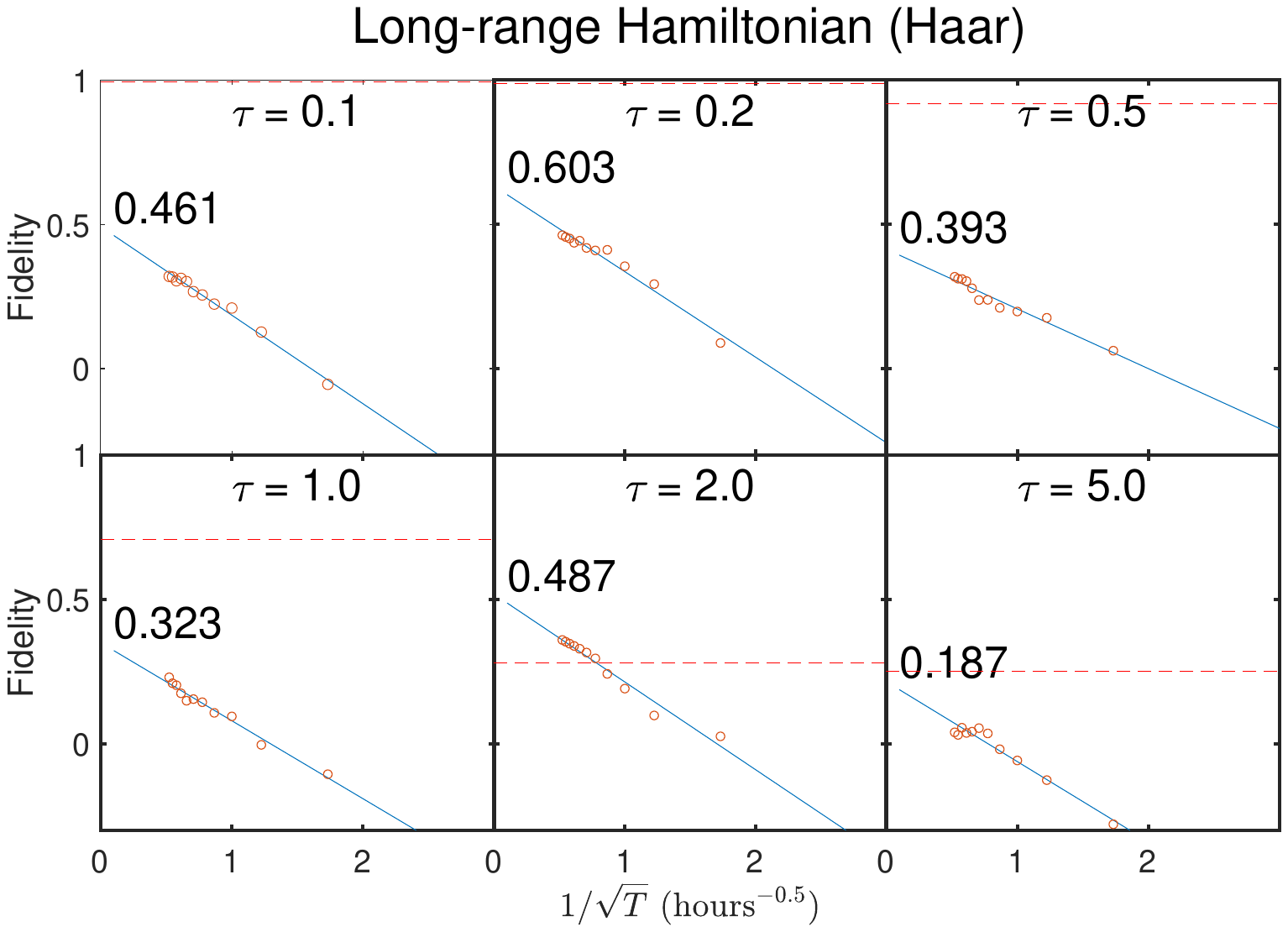} 
    \caption{\textbf{Convergence of certification statistics}. Certification of the first long-range order/Haar random 
    Hamiltonian system as a function of time for all 6 normalized time steps $\tau$, for confidence level $\epsilon = 0.9$. $T$ indicates the measurement time.
    }
    \label{fig:Haar_certification_linearized}
\end{figure}

\newpage
Similarly, the first long-range Haar-random system's certification converges as shown in Supplementary Fig.~\ref{fig:Haar_certification_linearized}. There are a total of $11$ batches for each time step. Here, the simulated time step of $\tau=2.0$ is clearly certified against the bi-partition. Unfortunately, the other time steps will not be able to reach the required certification fidelities when the measurement time is increased.

\end{widetext}

%	\newpage
\bibliographystyle{apsrev4-2}

%\bibliography{refsTherm_Selected.bib}

\begin{acknowledgements}
The Berlin team acknowledges funding from the BMBF (QPIC-1, PhoQuant), 
DFG (specifically EI 519/21-1 on paradigmatic quantum devices, but also CRC 183, project A03, on entangled states of matter, and FOR 2724 on quantum thermodynamics), the FQXi, the Einstein Foundation (Einstein Research Unit), 
the Munich Quantum Valley (K-8), and the Studienstiftung des Deutschen Volkes. It has also received funding from the European Union's Horizon 2020 research and innovation programme (PASQuanS, PASQuanS2). The Twente team acknowledges funding from the Nederlandse Organisatie voor Wetenschappelijk Onderzoek (NWO) via QuantERA QUOMPLEX (Grant No.~680.91.037) and Veni (grant No.~15872). J.~L. acknowledges funding from the European Research Council (QUCHIP, H2020-FETPROACT-2014). I.~A.~W. acknowledges funding from the European Research Council (Advanced Grant MOQUACINO), the Engineering and Physical Sciences Research Council (projects N509711, K034480, P510257 and T001062) and H2020 Marie-Sklodowska-Curie Actions (project 846073). \\
  
\end{acknowledgements}

\subsection*{Author contributions} J.~F.~F.~B., J.~L., I.~A.~W., N.~W., and J.~J.~R.~conceived the initial idea of this experiment and conducted initial simulations. H.~J.~S., M.~D.~G., B.~K., P.~V., C.~T., J.~P.~E., H.~H.~vdV., J.~T.~and
J.~J.~R.~constructed the photonic processor. F.~H.~B.~S., N.~W., J.~J.~R.~developed the experimental protocol. F.~H.~B.~S., R.~V.~D.~M., and M.~C.~A. performed the experiment and conducted data analysis under supervision of P.~H.~W.~P. and J.~J.~R. R.~S., J.~E.~and N.~W.~developed the verification scheme. F.~H.~B.~S., R.~V.~D.~M., M.~C.~A., R.~S., P.~W.~H.~P., J.~E.~, N.~W.~ and J.~J.~R.~wrote the manuscript, and all authors provided feedback on the text. J.~E.~and J.~J.~R.~provided overall guidance to the project. 

\subsection*{Competing interests} 
J.~J.~R.~and P.~W.~H.~P.~are shareholders in QuiX Quantum B.~V. 
The remaining authors declare no other competing interests.
%The other authors declare no competing interests.

\end{document}